\pdfoutput=1
\documentclass[tradiabstract]{aa} % for the abstract without structuration

\usepackage{graphicx,txfonts}
\usepackage{amsmath}
\usepackage{mathtools}
\usepackage{relsize} 
\newcommand{\simgreat} {\mathbin{\lower 3pt\hbox{$\rlap{\raise
5pt\hbox{$\char'076$}}\mathchar"7218$}}}

\newcommand{\simless}{\mathbin{\lower 3pt\hbox {$\rlap{\raise
5pt\hbox{$\char'074$}}\mathchar"7218$}}}

\begin{document}

\title{Dark dust~{\sc {II}}: Properties in the general field of the diffuse ISM}

\titlerunning{Dark dust in the diffuse ISM}

\author {R.~Siebenmorgen}
\institute{European Southern Observatory, Karl-Schwarzschild-Str. 2, D-85748
Garching b. M\"unchen, Germany; email: Ralf.Siebenmorgen@eso.org}

\date{Received April 25, 2022 / Accepted xxx, xxx}

\abstract{Distance estimates derived from spectroscopy or parallax
have been unified by considering extinction by large grains. The
addition of such a population of what is called {\it dark dust} to
models of the diffuse interstellar medium is tested against a
contemporary set of observational constraints. The dark dust model
explains, by respecting representative solid-phase element abundances,
simultaneously the typical wavelength-dependent reddening, extinction,
and emission of polarized and unpolarized light by interstellar dust
particles between far UV and millimetre wavelengths. The physical
properties of dark dust are derived. Dark dust consists of
micrometre-sized particles, which have been recently detected
in$-$situ. It provides significant wavelength-independent reddening
from the far UV to the near infrared. Light absorbed by dark dust is
re-emitted in the submillimeter region by grains at dust temperatures
of $8-12$\,K. Such very cold dust has been frequently observed in
external galaxies. Dark dust contributes to the polarisation at
$\simgreat \, 1$\,mm to $\sim 35$\,\% and at shorter wavelengths
marginally. Optical constants for silicate dust analogous are
investigated. By mixing 3\% in mass of Mg$_{0.8}$Fe$^{2+}_{0.2}$
SiO$_3$ to MgO$-$0.5 SiO$_2$ a good fit to the data is derived that
still can accommodate up to 5 - 10\% of mass in dark dust.  The extra
diming of light by dark dust is unexplored when discussing SN~Ia light
curves and in other research. Previous models that ignore dark dust do
not account for the unification of the distance
scales.}  \maketitle

%%%%%%%%%%%%%%%%%%%%%%%%%%%%%%%%%%%%%%%%%%%%%%%%%%

\section{Introduction \label{intro.sec}}

Physical properties of dust in the diffuse interstellar medium (ISM)
are derived by confronting models against observational constraints.
Two classes of models emerged over the years, either those that employ
distinct grain populations \citep{Mathis77, DraineLee84, Desert90,
WD01, Zubko04, Compiegne11, S14} or models which are using a mixture of
different grain constituents forming composite particles
\citep{Hage90, Mathis90, Ossenkopf91, KS94, GreenbergLi,V12,Jones13,
S14, Koehler15, Ysard15, Jones17, Guillet18, DH21}. The present
observational status of ISM dust characteristics after the Planck
mission \citep{Planck14} has been reviewed by \cite{HD21}. Dust models
shall agree with several major observations:

{\it a)} The solid-phase element abundances in the medium out of which
the dust is made \citep{HD21}.

{\it b)} The reddening curves in the Milky Way \citep{FM07, G09, F19}.

{\it c)} The diffuse galactic dust emission that has been observed in
the wavelength range from a few $\mu$m up to several mm by space
missions ISO, AKARI, Spitzer, WMAP, DIRBE \citep{Dwek97}, and Planck
\citep{Planck14}.

{\it d)} The optical/near-infrared (NIR) starlight polarization
\citep{Serkowski}, which is due to the dichroic extinction of aligned
non-spherical dust particles \citep{HG80, K08, Draine09, S14, V12}.

{\it e)} The polarized emission spectrum of the same grains which is
predominantly observed by \citet{Planck20}. Strikingly previous dust
models in the pre-Planck era systematically under-predict the observed
submillimeter (submm) and mm emission of unpolarized light from about
$0.3 - 3$\,mm, and did not explain the flatness of the polarized dust
emission spectrum in that wavelength range.

\noindent
{\it f)} In addition, the unification of spectroscopically derived
distances with the parallax is not closed.

However, previous dust models failed to explain all observational
constraints that have been available in the post-Planck era
simultaneously \citep{Ysard20}, unless the model by \citet{DH21}, in
which the optical constants of amorphous silicates were modified to
fit the Planck data.

For a star the spectro-photometric distance $D_{\rm{SpL}}$ shall agree
with the distance $D_{\rm {GAIA}}$ derived from parallax. Geometric
distances using GAIA \citep{Prusti} as derived by \citet{Bailer18,
Bailer21} show for OB stars with available reddening curves at
distances below 2\,kpc a fractional error of $\sim 5$\,\%. The
photometric distance of a star is connected via the apparent $m_{\rm
V}$ and absolute magnitude $M_{\rm V}$, and the dust extinction
$A_V$ along that sightline: {$\log{D_{\rm{SpL}}} = 0.2 \ (m_V - M_V -
A_V + 5)$\,pc}. Photometric distances require accurate calibration
of the spectral type and luminosity class (SpL) of the star allowing
derivation of $M_V$ and $A_V$ (Eq.\ref{RvObs.eq}). The original SpL
estimates of OB stars \citep{W71, WF90} have gained significantly in
precision using results of the galactic O star spectroscopic survey
\citep{Maiz04} with SpL updates by \citet{Sota11, Sota14} and applying
quantitative spectral classification schemes utilizing high-resolution
optical spectroscopy \citep{Markova11, Martins18}. Standard grids of
absolute magnitudes for OB stars are provided
\citep{Bowen08,Pecault13}\footnote{http://www.pas.rochester.edu/\~emamajek/}
as well as high-quality extinction estimates \citep{ Maiz18}. This
progress enables computing spectro-photometric distances of OB stars
at fractional precision of $\sim 15\%$ for the nearby (2\,kpc) sample.
Photometric versus parallax distance estimates to galactic OB stars
have been discussed by \citet{Shull19} and reveal larger uncertainties
at larger distances. In the nearby sample, the spectroscopic distances
show a systematic overestimate with a dispersion above these errors
\citep{S20}. Both distance estimates of the same source shall of
course agree.

Already \citet{Trumpler} included in the original form of the
photometric equation, in addition to the wavelength-dependent
(selective) interstellar extinction, a constant,
wavelength-independent (non-selective) extinction component in the
optical. He proposed that light may be obscured by large
('meteoritic') particles. Large micrometre-sized dust particles are
frequently found in circumstellar shells \citep{Strom, Jones72,
Lanz95, Steinacker15, Kataoka17}, Herbig stars \citep{Dunkin98},
$\eta$ Car \citep{Andriesse78} and other evolved stars
\citep{Scicluna15, Scicluna22}. Grains larger than $2 \, \mu$m have
been seen in scattering light halos around X-ray sources
\citep{Witt01} and grains as large as $\sim 4 \, \mu$m have been
suggested for V404 Cygni by \citet{Heinz16}. The emission from
100\,$\mu$m large particles accounting for the observed submm fluxes
from evolved giants are derived by \citet{Jura01}. \citet{Maercker22}
finds that for carbon stars preferentially $\sim \, 2\mu$m sized
grains survive the AGB wind-ISM interaction regions and act later as
seeds for grain growth in the ISM.

Distance estimates of the Orion Trapezium star HD~37020 using spectral
type-luminosity distance $D_{\rm SpL}$ are a factor 2.5 larger than
the Ca\,{\sc ii} or VLBI parallax distance estimates
\citep{Krelowski16}. For several nearby ($\la 400$\,pc) OB stars a
significant overestimate of the spectroscopic distance over the
HIPPARCOS parallax is reported by \cite{Skorzynski03}. Assuming that
the parallax has been correctly measured the derived absolute
magnitudes of these stars appear too faint. ''Super large'' grains are
suggested for the extra weakening of the observed brightness of these
stars. The column density of this very large grain population is well
correlated with the strength of {DIB~6367\,~\AA}\, and
{DIB~6425\,~\AA}\, and therefore appears distributed along the
sightline through the diffuse ISM. The spectral type-luminosity
distances of 132 OB stars have been compared to those derived by GAIA
up to $D_{\rm GAIA} < 2$\,kpc \citep{S20} and for $\sim 10$\% of the
sample $D_{\rm {SpL}} / D_{\rm {GAIA}} \simgreat 2$. Both distance
estimates have been unified introducing $0.3-0.7$\,mag reddening by
what they called {\it dark dust}. The terminology follows the
unveiling of {\it {dark gas}} by \citet{Grenier05} because gas and
dust are intimately mixed in the diffuse ISM. It describes a
mysterious, dimmed and hidden nebulosity. Extensive clouds of dark gas
have been weighted using Fermi-LAT $\gamma$-ray data
by \citet{Widmark22} and emerge in the cold ISM. The hidden dust
component appears in sightlines that are connected to the cold ISM as
well and have line intensity ratio of CH 4300\,\AA\, to CH$^+$
4233\,\AA \ greater than one and CN 3875\,\AA\ detected. Possibly a
sticking of smaller grains into larger units is favoured in cold ISM
environments. A circumstellar nature of dark dust towards these stars
is excluded from inspecting WISE imaging between $3 -
22 \, \mu$m. Micrometre-sized particles from the diffuse ISM have been
measured {\it in$-$situ} from the Ulysses, Galileo, and Stardust space
probes to the outer solar system \citep{Landgraf, Westphal14,
Krueger15}.

The emission of dark dust as a new component of the diffuse ISM should
be observable in the submm/mm wavelength range. It will absorb a
fraction of the interstellar radiation field, ISRF
\citep{Mathis83}. As such grains are big they are cold and will emit
at long wavelengths. Originally very cold ($\la 15$\,K) dust emission
has been detected in the general field of the ISM of non-active
galaxies \citep{KS98,S99} and in our Galaxy towards high-density
regions \citep{Chini93}. More recently, the Herschel KINGFISH
\citep{Kennicutt11} and Dwarf Galaxy \citep{Madden13} surveys find
that 35 out of 78 galaxies have excess emission at 0.5\,mm that cannot
be explained by a single modified black-body temperature component
with dust emissivity spectral index of two \citep{RemyRuyer13}. Other
galaxies, e.g. Haro~11 \citep{Galliano05,Galametz09}, show the excess
emission at even longer wavelengths in the mm range and were missed by
Herschel surveys. The excess emission is due to very cold dust at
temperatures as low as 10\,K. A solution for explaining the submm
excess emission is often favoured by ad-hoc adjustments of the
emissivity law of the grains \citep{RR13, Guillet18}. This can be
derived by changes in the porosity, shape, or adjustments of the
optical constants \citep{DH21}. Models changing the spectral index of
the dust emissivity do not touch on the issue of distance unification.

In this paper, a model for the general field in the diffuse ISM is
described that accounts simultaneously for observations of solid-phase
elemental abundances, average Milky Way reddening, IR to mm emission
at high galactic latitudes, average Milky Way polarized extinction and
polarized emission by dust. A particular feature is the inclusion of
dark dust as an additional dust population that accounts for the very
cold emission that is detected in the submm/mm range. Dark dust may
solve the puzzle between discrepant spectroscopic and parallax-derived
distance estimates. First, the observational basis is specified, then
the dust model is described, and a vectorized fitting procedure of the
11-parameter dust model to the observational constraints is
detailed. Results will be presented that show the overall success of
the model in fitting the data and deriving the physical properties of
dark dust. Particular attention is given to the use of optical
constants of amorphous silicates \citep{Demyk22}. The main findings
are summarized in the conclusions.

\section{Observations \label{obs.sec}}
\subsection{Soild-phase abundances}

Depletion of elements from the gas into dust is estimated by
\citet{VH10} as the difference between abundances in the Sun
\citep{Asplund09} and that of the gas \citep{Jenkins09}. Elemental
depletion is used to infer dust compositions and some form of grains
with silicate and carbon is widely accepted. Abundances of element [X]
to that of hydrogen [H] are summarized by \citet{HD21} for the gas and
the dust phase, respectively. The most abundant elements in dust are:
O ($249 \pm 94$), C ($126 \pm 56$), Mg ($46 \pm 5$), Fe ($43 \pm 4$),
Si ($38 \pm 3$), S (7.6), Al (3.4), Ca (3.2), and Ni (2.0), where
numbers in parenthesis give [X]/[H] (ppm). An over-abundance of O in
the dust is noted and upper limits of the $16\,\mu$m and $22\,\mu$m
bands of iron-oxides indicate that most of the Fe remain unexplained
as well \citep{DH21}. However, the Fe abundance is insufficient to
form large grains \citep{Zhukovska18}. Fe particles heated by the ISRF
contribute to the emission at $\sim
60\,\mu$m \citep{Fischera04}. \citet{DH21} showed that the absorption
cross-section remains in the IR - mm unaltered when including the
available Fe in form of impurities in large particles.

Dust abundances are uncertain and estimates of
the C and Si abundance in dust scatter by 50\%. \citet{Mulas13} find
an average of [C]/[H] $=145$\,ppm. The Si abundance in the dust is
estimated by \citet{Sofia01} to be $18 \pm 9$\,ppm, while
\citet{Nieva12} derive $32 \pm 3$\,ppm. \citet{VH10} find a
sharp difference in the dust abundances for sightlines located at low
and high galactic latitudes and give an average abundance of [Si]/[H]
$= 23 \pm 5$\,ppm.

Absorption and emission signatures of the dust provide important
constraints on the composition. A striking feature is the
2175\,\AA\, extinction bump, where graphite and polycyclic aromatic
hydrocarbons (PAHs; \citet{Allamandola89}) have strong electronic
transitions. In the IR, there are conspicuous emission bands at 3.3,
6.2, 7.7, 8.6, 11.3, and 12.7\,$\mu$m, as well as a wealth of weaker
bands in the $12 - 24\,\mu$m region. These bands are ascribed to
vibrational transitions in PAH molecules, which are planar structures
that consist of benzol rings with hydrogen attached. PAH feature
strengths depend on the hardness of the exciting radiation field and
the ionisation or hydrogenation coverage of the molecules.

Absorption features in the diffuse ISM between $3 - 8 \, \mu$m are
interpreted as being due to either carbonaceous material
\citep{Mennella03} or, guided by laboratory spectra, by ices mixed
with silicates \citep{Potapov21}. Dust models including interstellar
ice mixtures have been presented by \citet{S97}. Solid-state water
mixed with silicates would explain a good fraction of the unaccounted
oxygen depletion in the diffuse ISM \citep{Potapov21}. A comparison of
future JWST observations against laboratory spectra is needed to
confirm the existence of solid water in the diffuse ISM. Ices are not
further considered in this study.

The 9.7\,$\mu$m and 18\,$\mu$m broadband features are assigned to Si-O
stretching and O-Si-O bending modes of silicate grains,
respectively. A comparison of the observed band profiles and
laboratory spectra favours amorphous rather than crystalline
silicates. Intriguing is the detection of the $11.1\,\mu$m
absorption band that is attributed in the atlas of ground-based
mid-infrared (MIR) spectra by \cite{Do20} to forsterite, providing a
mass abundance of crystalline silicates of $\sim 2$\% in the diffuse
ISM. Olivine (Mg$_{2x}$ Fe$_{2-2x}$ SiO$_{4}$ has $x \sim 0.8$),
have been detected in AGB and T Tauri stars, and comet Hale-Bopp
\citep{H10}. The silicate stoichiometry and grain geometry
can be revealed by MIR spectro-polarimetry (Sec.~\ref{polobs.sec}).
As nominal composition \cite{DH21} adopt
Mg$_{1.3}$(Fe,Ni)$_{0.3}$SiO$_{3.6}$ having a molecular weight of
$\mu = 134.5$. In that structure, the silicate grain abundance is
limited by Mg with a dust abundance ratio $\rm{[C]/[Si]} \sim
\rm{[C] \ 1.3/[Mg]} \sim 3.5$. To accommodate the reported large
scatter of $\sim 50$\,\% in the solid-phase element abundances the
dust models shall respect a limit of the dust abundance ratio of

\begin{equation} \label{abu.eq}
\rm{[C]/[Si]} \la 5.25 \ .
\end{equation}

\subsection{Reddening and extinction}
\label{reddening.sec}

The interstellar reddening is derived by measuring the flux ratio of a
reddened and unreddened star with the same SpL
luminosity class, e.g. the standard pair method
\citep{Stecher}. The flux of a star is derived from the spectral
luminosity $L(\lambda)$, the distance $D$, and the extinction optical
depth $\tau(\lambda)$, which is due to the absorption and scattering of
photons along the sightline excluding emission. The observed flux of
the reddened star is given by

\begin{equation} \label{flux.eq}
F(\lambda) = \frac{L({\lambda}) }{ 4 \pi \ D^2} \ e^{-\tau({\lambda})} \ .
\end{equation}

\noindent In photometry it is customary to express the flux of an
object by the apparent magnitude, which is related to the flux through
$ m(\lambda) = 2.5 \log _{10} \Big( F(\lambda)/w_{\lambda} \Big) $ and
$w_{\lambda}$ as the zero point of the photometric system. The difference
in magnitudes between two stars is $\Delta m(\lambda) = 1.086 \times
\Big( \tau(\lambda) + 2 \log _{10}(D/D_0) \Big ) $\,. Unfortunately,
distances to hot, early-type stars, commonly used to measure
interstellar extinction, are often subject to large errors
\citep{S20}. Hence one relies on relative measurements of two
wavelengths and defines the colour excess {$ E(\lambda -
\lambda')=\Delta m(\lambda)-\Delta m(\lambda')$}. The {\it
{reddening curve}} $E(\lambda)$ is traditionally represented by a
colour excess and normalisation to avoid the uncertainties between
both distances,

\begin{equation} \label{eq1}
E(\lambda)= \frac {E (\lambda - \rm {V})} {E(\rm{B}-\rm{V})} = \frac
{A_{\lambda} -A_{\rm {V}} } { A_{\rm {B}} - A_{\rm {V}}} = \frac
{\tau_{\lambda} - \tau_{\rm{V}} } { \tau_{\rm {B}} - \tau_{\rm {V}}} \ .
\end{equation}

\noindent Naturally, for the V and B band, $E(\rm{V}) = 0$ and
$E(\rm{B}) = 1$. The extinction in magnitudes at wavelength $\lambda$
is denoted by $A(\lambda)$. One obtains some extrapolated estimate of
the visual extinction $A_{\rm {V}}$ from photometry. This requires
measuring $E(\rm{B}-\rm{V})$ and extrapolating the reddening curve to
infinite wavelengths $E(\infty)$. In practice, one derives
$E(\lambda-\rm {V})$ at the longest wavelength, which is not
contaminated by any kind of emission. From this wavelength, e.g. the
H-band, one extrapolates to infinite wavelengths assuming some prior
shape of $E(\lambda)$ and hence estimates $E(\infty,V)$. By
introducing the ratio of the {\it {total-to-selective extinction}}

\begin{equation} \label{RvObs.eq}
R_{\rm V} = \frac{A_{\rm V}} {E(\rm{B} - \rm{V})} \ , 
\end{equation}
\noindent
where obviously $E(\infty) = - {R_{\rm V}}$. The total-to-selective
extinction can also be written using Eq.~\ref{eq1} or applying a dust
model as

\begin{equation} \label{Rv.eq}
{R_{\rm V}} =\frac{\tau_{\rm{V}} } { \tau_{\rm {B}} - \tau_{\rm {V}}} = \frac {K_{\rm{V}} } {K_{\rm {B}} - K_{\rm {V}}} \ , 
\end{equation}

\noindent where $K = K_{\rm {abs}} + K_{\rm {sca}}$ is the extinction
cross-section which is the sum of the total absorption and scattering
cross-section of the dust. The relation between reddening and
extinction is
\begin{equation} \label{tau.eq}
\frac {\tau(\lambda)} {\tau_{\rm V}} = \frac {K(\lambda)} {K_{\rm V}} = \frac{E(\lambda)} {R_{\rm V}} + 1 \ ,
\end{equation}

\begin{figure*}[htb]
\begin{center}
\includegraphics[width=9cm,clip=true,trim=4.cm 7.cm 4.cm 7.cm]{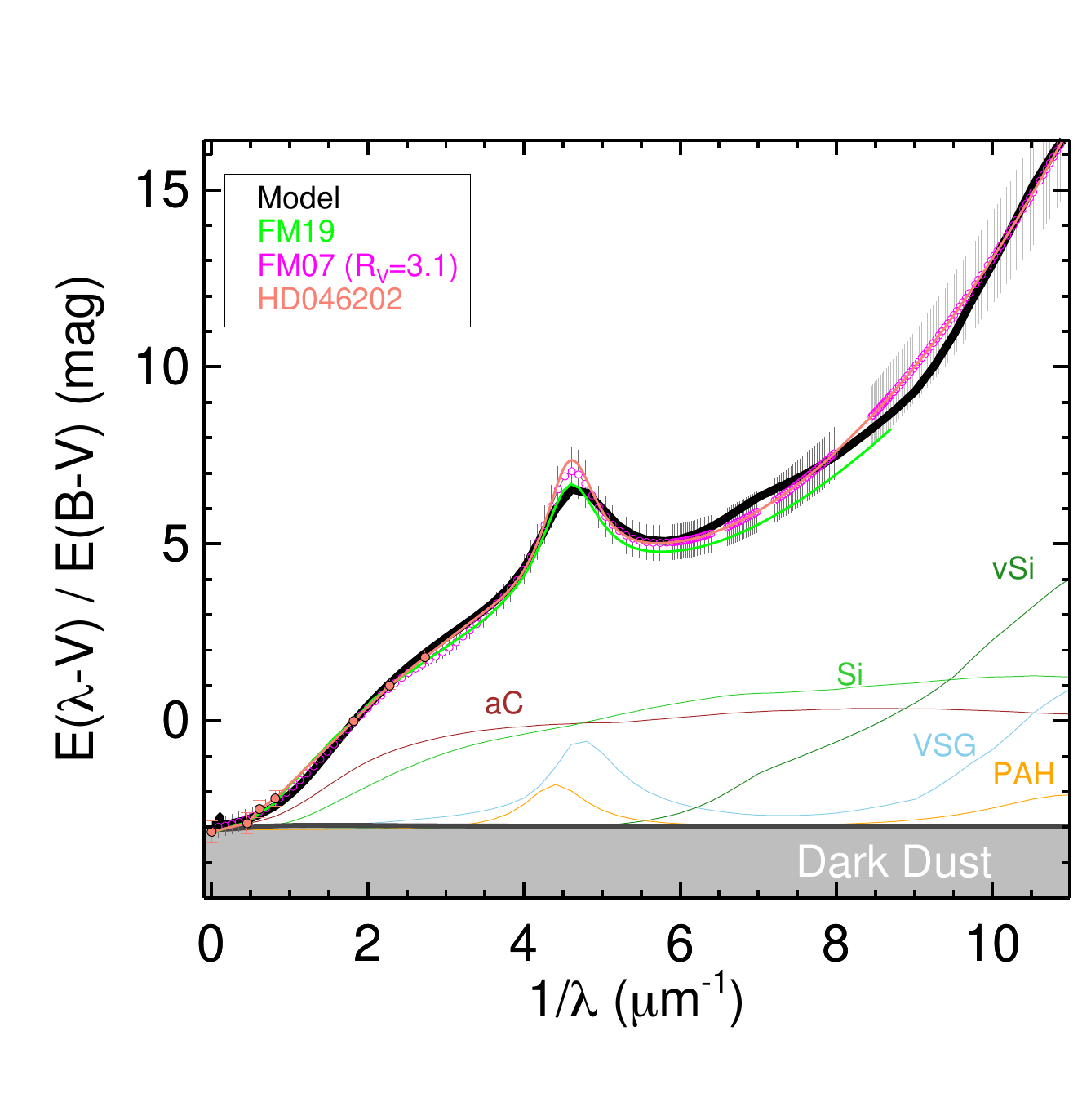}
\includegraphics[width=9cm,clip=true,trim=3.7cm 6.7cm 3.8cm 6.8cm]{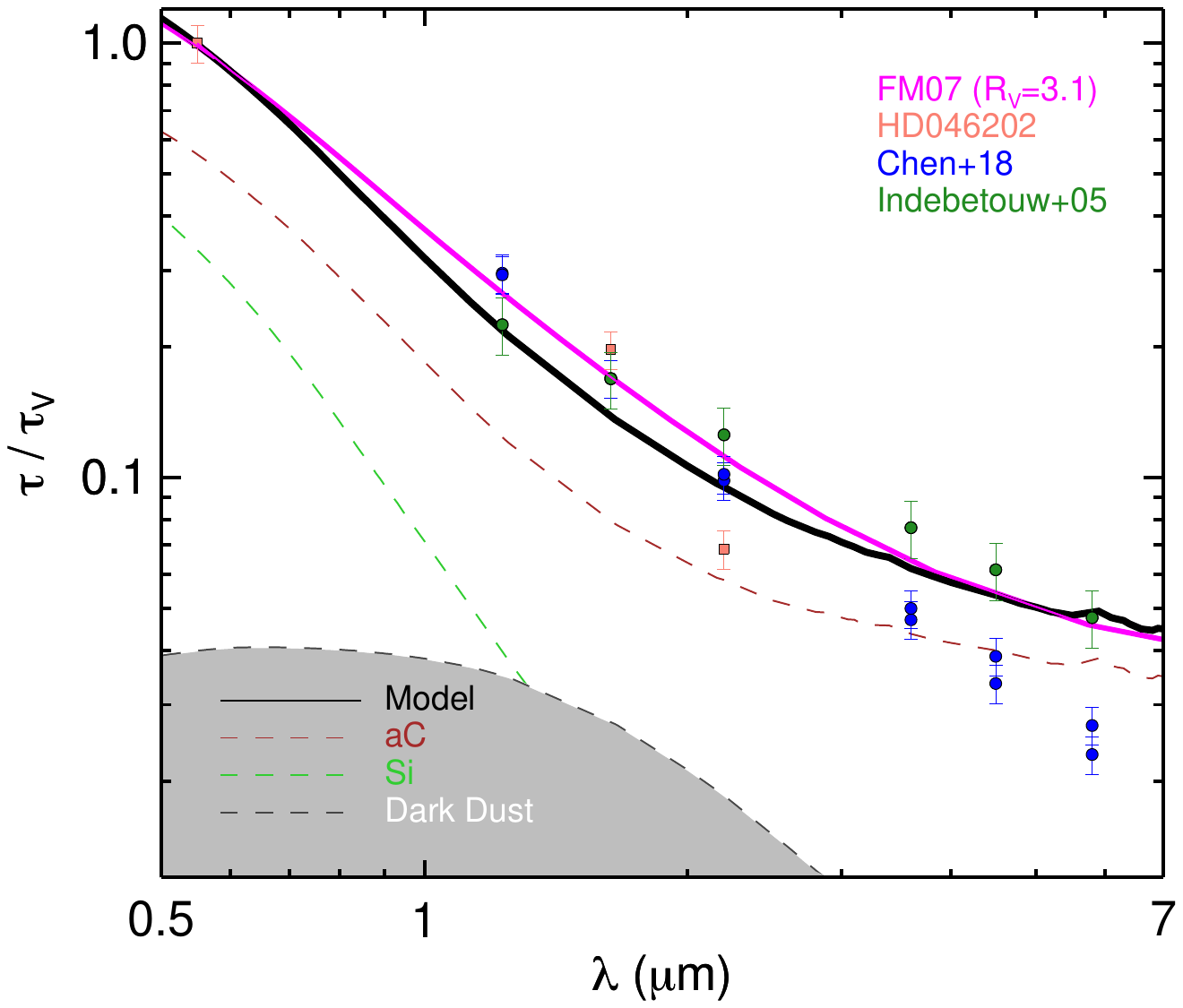}
\end{center}
\caption {Reddening (left) and extinction (right) of the diffuse ISM
  \citep{FM07, F19, Indebetouw05, Chen18} and of
  HD~046202 \citep{G09}. Model (black) with $r_{\rm {Dark}}^{+} =
  1 \, \mu$m (Table~\ref{para.tab}) and individual dust components as
  labelled. The area in grey shows the contribution of dark
  dust. \label{redd.fig}}
\end{figure*}

The International Ultraviolet Explorer (IUE) operated between $0.185 -
0.33\,\mu$m and the Far Ultraviolet Spectroscopic Explorer (FUSE)
observed between $0.119 - 0.905\,\mu$m. IUE and FUSE provide the
legacy of (far) UV spectra with a total of 895 reddening curves
towards more than 568 early-type (OB) stars
\citep{Val04,FM07,G09}. The database of available reddening curves was
scrutinized against systematic errors by the following means: 1)
IUE/FUSE spectra were verified to include only a single star in the
observing aperture and not beside the program star and other almost
equally bright objects. 2) Although most early-type stars are binaries
\citep{C12} their reddening curves are generally derived assuming a
single star. Whenever the companion contributes significantly to the
total flux of the system, the derived extinction includes large
systematic errors. 3) A large fraction of the OB stars show emission
components above the photosphere in the IR \citep{S18b,Deng22}, which
prohibits deriving meaningful reddening in the IR. 4) Reddening curves
are derived assuming steady stellar systems. Stellar variability will
thus systematically impact the derived reddening. Stars with detected
variability between ground-based and Hipparcos V-band photometry ($\ga
0.1$\,mag), significant B - V colour changes, or GAIA G-band
photometry of more than ($\ga 0.033$\,mag) shall be rejected. 5) some
stars show inconsistencies in the GAIA parallaxes between data release
two \citep{DR2} and data release three \citep{DR3} that are not
instrumental so that their reddening curves shall be declared as
spurious. 6) Finally, the quality of the derived reddening curve
depends critically on the fidelity of the SpL estimate. The SpL
determination of fast rotators is highly uncertain and UV spectral
diagnostics indicate considerably earlier SpL (hotter) classifications
than optically assigned SpL. Reddening curves with largely deviating
SpL assignments shall be ignored. All these systematic effects impact
the derivation more dramatic at $\lambda \simgreat 1 \, \mu$m and the
extinction more than the reddening. In total, 48 stars with one or
more reddening curves passing the rejection criteria are available
with 21 classified as multi-component sightlines and 27 as
single-cloud sightlines \citep{S20}. The reddening curves in the ISM
show significant variations from sightline-to-sightline and the
derived total-to-selective extinction are between $2 \la R_{\rm V} \la
6.4$. For the same star the published $R_{\rm V}$ that are mostly
extrapolated from JHK photometry, scatter typically by 10\%. Even in
the high-quality sample, the peak-to-peak scatter in $ R_{\rm V}$
estimates of the same star are up to $\pm 0.6$. The scatter might be
reduced by detailed physical modelling of the dust applying
Eq.~(\ref{Rv.eq}). Whenever possible the reddening instead of the
extinction curve shall be discussed.

For the diffuse ISM a mean reddening of ${E(\rm{B} - \rm{V})} =
0.45$\,mag and a median and mean value of $R_{\rm V}= 3.22$ and $2.99
\pm 0.27$ is derived by \citet{FM07}, respectively; \citet{Wang19}
find $R_{\rm V}= 3.16$, and \citet{V04,F19} give ${A_{\rm V}} \sim 3.1
\ E({B-V)}$. Here $R_{\rm V} = 3.1$ is adopted as by \cite{HD21}. The
mean reddening curve of the Milky Way is derived for $1/\lambda \la
1./8\,\mu$m by \citet{FM07} and \citet{F19} and both are shown
together with the reddening curve of HD~046202 \citep{G09} in
Fig.~\ref{redd.fig}. The reddening curve of that star perfectly
matches the mean curve derived from IUE and FUSE. Reddening in spectral
regions close to wind lines at 6.5 and 7.1\,$\mu$m$^{-1}$ and
Ly-$\alpha$ at $8\,\mu$m$^{-1} \leq x \leq 8.45\, \mu$m$^{-1}$, or
with apparent instrumental noise at $ x \simless 3.6\,\mu$m$^{-1}$
shall be ignored. Overall a typical error in the derived reddening is
$\sim 10$\,\%. Various mean extinction curves of the diffuse ISM
between $0.5 - 7\,\mu$m are also shown in Fig.~\ref{redd.fig}.
Noticeable is the large scatter which is increasing towards longer wavelengths.

%%%%%%%%%%%%%%%%%%%%%%%%%%%%%%%%%%%%%%%%%%%%%%%%%%
\begin{figure}[htb]
\includegraphics[width=9.cm,clip=true,trim=4.cm 8.cm 3.5cm 8cm]{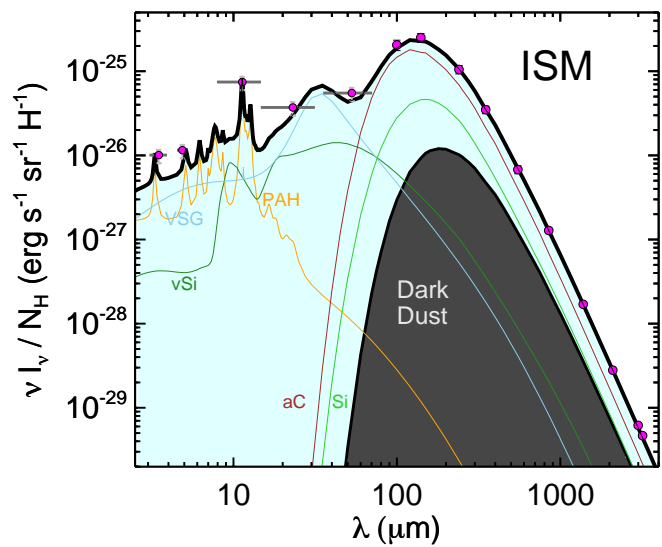}
\caption {Dust emission of the diffuse ISM observed at high galactic
  latitudes and normalised per H atom by \citet{Dwek97}
  and \citet{Planck15} and as tabulated by \citet{HD21}. Model (black)
  with $r_{\rm {Dark}}^{+} = 1 \, \mu$m (Table~\ref{para.tab}) and
  individual dust components as labelled. The area in grey shows the
  contribution of dark dust. \label{sed.fig}}
\end{figure}

%%%%%%%%%%%%%%%%%%%%%%%%%%%%%%%%%%%%%%%%%%%%%%%%%%

\subsection{Diffuse galactic dust emission}

Diffuse emission of the ISM from dust grains heated by the ambient
ISRF has been observed from the NIR through the microwave region. The
dust emission of the diffuse ISM is not uniform across the sky and
there is evidence that the properties of the dust as well as the
spectral distribution and strength of the ISRF that heats the dust
vary \citep{Fanciullo15}. Observations of this component for high
galactic latitude ($\vert b \vert \ga 25\degr$) from the Cosmic
Background Explorer using DIRBE data are given by \citet{Dwek97} and
from the \citet{Planck15}. A colour-corrected composite spectrum with
error estimates is tabulated by \citet{HD21}. The emission spectrum is
shown in Fig.~\ref{sed.fig}.

\subsection{Starlight and FIR polarization \label{polobs.sec}}

Stellar light that passes a cloud of moderate extinction by aligned
grains becomes linearly polarized. From the far UV to the NIR, between
$0.12\,\mu\rm{m} - 1.6\,\mu$m, the polarization curves are fit by an
empirical formula given by \citet{Serkowski}

\begin{equation}
{p(\lambda)} = {p_{\max}} \ \exp \left[ -k_{\rm p} \ \ln^2
\left( \frac{\lambda_{\max}}{\lambda} \right) \right]\,.
\label{Serk.eq}
\end{equation}

\begin{figure}[htb]
\includegraphics[width=9cm,clip=true,trim=4.cm 7.cm 3.5cm 8.cm]{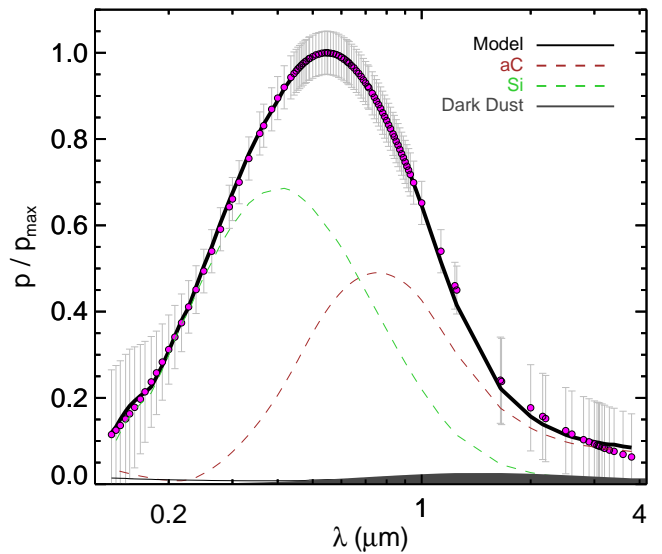}
\caption {Polarized extinction of the diffuse ISM (circles) with error
  bars indicating the observed scatter \citep{VH12,Bagnulo17}. The
  model (black) with $r_{\rm {Dark}}^{+} = 1 \, \mu$m
  (Table~\ref{para.tab}) and contributions (dashed) from large
  amorphous carbon (brown), large silicate grains (green), and dark
  dust area in grey, when treated as prolate particles, is
  shown. \label{Serk.fig}}
\end{figure}
\begin{figure}[htb]
\includegraphics[width=9cm,clip=true,trim=4.cm 7.cm 3.5cm 8.cm]{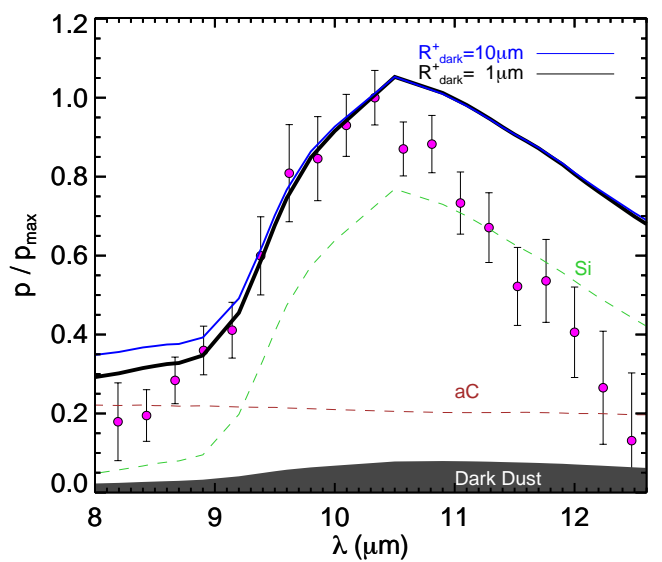}
\caption {Polarized extinction in the MIR. A composite spectrum
(circles) of two Wolf-Rayet stars WR~48A and AFGL~2104 by
\citet{Wright02}, the model (full line) with $r_{\rm {Dark}}^{+} = 1
\, \mu$m (black) and contribution (dashed) from large amorphous
carbon (brown), large silicate grains (green), and dark dust
(grey) when treated as prolate particles, is shown, as well as
the model with $r_{\rm {Dark}}^{+} = 10 \, \mu$m (blue). Parameters
as in Table~\ref{para.tab}. \label{mirpol.fig}}
\end{figure}


%%%%%%%%%%%%%%%%%%%%%%%%%%%%%

\noindent
where $\lambda_{\max}$ is the wavelength at maximum polarization
$p_{\max}$ and $k_{\rm p}$ the width of the spectrum. Significant
variations in the width $0.5 \la k_{\rm p} \la 1.5$ towards different
sightlines are observed, using HPOL{\footnote
{http://www.sal.wisc.edu}}, the Wisconsin UV Photo-Polarimeter WUPPE
satellites, and ground-based instruments as compiled by
\citet{Efimov09} and by the Large Interstellar Polarization Survey
\citep{Bagnulo17}. The latter authors could not confirm the linear
trend of $k_{\rm p} \propto \lambda_{\rm max}$ claimed in earlier work
\citep{Wilking80, Whittet92}. Observations of 160 sightlines of
mildly reddened stars show that $p_{\rm{V}} /E_{B-V} \la 9$\,\%/mag
and $p_{\rm {max}} \la 10$\,\% \citep{V16}. Starlight polarization
reaches a maximum in the V band at $\lambda_{\rm max} =0.545\,\mu$m
and $k_{\rm p} = 1.15$ is selected to represent the mean Serkowski
curve that shows a typical scatter of 5\% near $p_{\rm {max}}$, 10\%
at $p/p_{\rm {max}} \la 0.4$, and in the NIR $\sim 15$\,\%. At
$\lambda > 1.5\,\mu$m the polarization spectrum smoothly matches onto
a power-law with an exponent of 1.6 \citep{Martin92}. The fit naturally
breaks in the MIR near the silicate band. The typical wavelength
dependence of the observed polarized extinction is shown in the
optical/NIR in Fig.~\ref{Serk.fig} and in the MIR in
Fig.~\ref{mirpol.fig}.

A mean MIR polarized extinction spectrum, which covers the silicate
band, is constructed by \citet{Wright02} by averaging observations of
two Wolf-Rayet stars, WR~48A and AFGL~2104. The analysis by Wright
shows that the polarized extinction of silicates with stoichiometry
and optical constants $(n, k)$ of Mg$_{0.4}$Fe$_{0.6}$SiO$_{3}$,
MgFeSiO$_{3}$, as well as MgFeSiO$_4$ provide a pure fit whereas
Mg$_{0.8}$Fe$_{1.2}$SiO$_{4}$ give an almost perfect fit to the
polarization spectrum \citep{Wright02}. The molecular weight of these
materials varies between $\mu = 97 - 141$, an uncertainty directly
impacting estimates of the gas-to-dust mass ratio. The silicate
composition observed towards the Wolf-Rayet stars is distinct from
Mg$_{1.3}$(Fe,Ni)$_{0.3}$SiO$_{3.6}$ that is adopted for the diffuse
ISM \citep{DH21}. There are marked differences between the
environments of the dust-producing circumstellar shells of the bright
Wolf-Rayet stars and the diffuse ISM. Towards both sightlines, the
stars have a factor $\sim 3$ higher reddening than what is typical for
the diffuse ISM and indications of ice absorption at $6.1\,\mu$m are
found \citep{MarchenkoMoffat17}. Therefore, the silicate composition
of the diffuse ISM is retained. Nevertheless, the analysis exemplifies
that characterization of the silicate stoichiometry and particle shape
by MIR spectro-polarization is viable. Unfortunately, such
observations were recorded in the past century \citep{Smith00} and
novel high-sensitive instrumentation is needed for detecting polarized
MIR extinction in the diffuse ISM.

The polarized dust emission spectrum in the submm is observed by
\cite{Planck14,Planck20}. If the polarized emission in the
submm and the polarized extinction in the optical is due to
the same grains then this ratio in principle includes information on
the grain elongation \citep{K03}. For the ratio of 850\,$\mu$m to V
band polarization a characteristic value of

\begin{equation}
\tau_{\rm {V}} \ \frac{p_{850\,\mu{\rm m}}}{p_{\rm {V}}} = 4.31 
\label{Pratio.eq}
\end{equation}

\noindent is adopted by \citet{HD21}. In the diffuse ISM variations
for different sightlines are noted in the hydrogen column density
$N_{\rm {H}}$, $E(\rm{B}-\rm{V})$, $R_{\rm V}$, and hence $\tau_{\rm
{V}}$ (Eq.\ref{tau.eq}).

\section{Model}

\subsection{Dust populations}

A dust model for the diffuse ISM is presented that shall be in line with
present observational constraints: the [C]/[Si] abundance ratio
(Eq.~\ref{abu.eq}), the spectral variation of the reddening, the
starlight polarization, the diffuse galactic emission, and the
polarized dust emission (Sec.~\ref{intro.sec}). Two major grain
materials are considered: amorphous silicates and carbon. Dust
particles need to be of different sizes to fit the reddening and polarization curves. A power-law size distribution
$\mathrm{d}n(r)/\mathrm{d}r \propto r^{-q}$ \citep{Mathis77} that
ranges from the molecular domain ($r_{-} \sim 5$ \AA\,) to a rather
unconstrained upper size limit of several microns ($r_{+} \la
10\,\mu$m) is applied. Three different dust populations are
distinguished:

{\it {1) Nanoparticles}} with sizes below 6\,nm, which are in form of
very small silicates (vSi), very small graphite (VSG), and PAHs. Two
kinds of PAH are treated, small molecules having 30 C and 12 H atoms
and clusters with 200 C and 40 H atoms, respectively. The
cross-sections of the nanoparticles and PAHs are taken from
\cite{S14}.

{\it {2) Large grains}} with radii between $6 - 300$\,nm, which are
partly aligned and of prolate shape with an axial ratio $a/b = 2$.
Large grains are made of silicates (Si) and amorphous carbon
(aC). Scattering, absorption, and polarization cross-sections of
spheroids are computed with the procedure outlined in
Sec.~\ref{spheroids.sec}.

%%%%%%%%%%%%%%%%%%%%%%%%%%%%%%%%%%%%%%%%%%%%%%%%%%
\begin{figure}[htb]
\includegraphics[width=9cm,clip=true,trim=2.5cm 4.3cm 3.5cm 5.5cm]{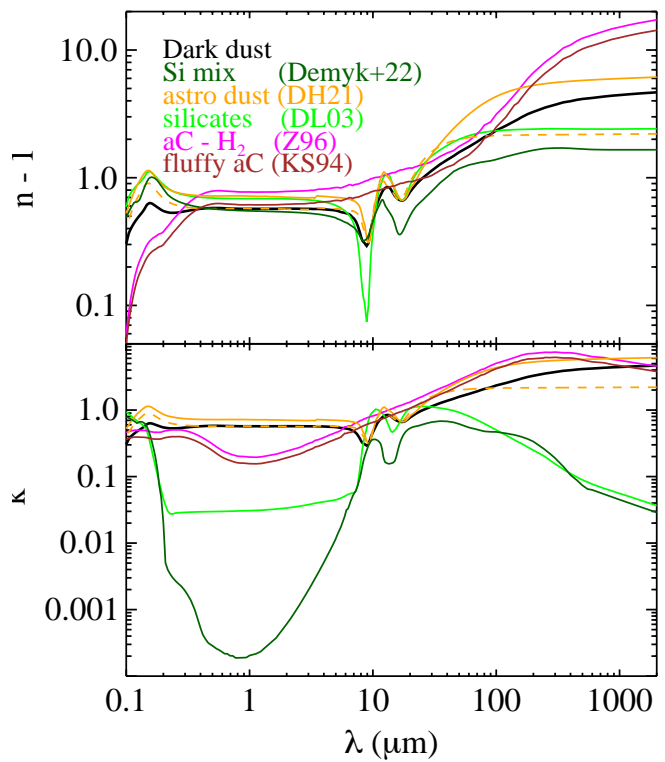}
\caption {Optical constants $n-1$ (top) and $k$ (bottom) for various
grain materials. \label{compnk.fig}}
\end{figure}

%%%%%%%%%%%%%%%%%%%%%%%%%%%%%%%%%%%%%%%%%%%%%%%%%%

{\it {3) Dark dust}}, is included as a new additional grain component
of micro-meter-sized particles. Dark dust grains are taken to be some
kind of fluffy aggregates that are made up of porous composites of
large silicate and large amorphous carbon grains that are sticking
loosely together. The frosting of molecules on dark dust can be
envisioned as an attractive reservoir of the reported O depletion but
is not considered because of the missing detection of ice absorption
bands in the diffuse ISM. Dark dust particles are treated as spheres
unless otherwise stated. For fluffy grains, there are only approximate
methods for calculating the cross-sections. The numerical simple tool
by \cite{KS94} is used utilizing the Bruggemann mixing rule. An
averaged complex dielectric function, $\epsilon = m^2$ is computed by
solving

\begin{equation}
\sum_{\rm i} f_{\rm i} \ {\frac{\epsilon_{\rm i} - \epsilon_{\rm
{av}}} {\epsilon_{\rm i} + 2 \epsilon_{\rm {av}}}} = 0 \ \quad
\ {\rm {with}} \quad \ \sum_{\rm i} f_{\rm i} =1 \ ,
\label{Brugg.eq}
\end{equation}

\noindent
where the volume fraction of each component $i$ is denoted by $f_{\rm
i}$. The considered inclusions in a dark dust particle are:
silicates with $f_{\rm Si}=0.5$, amorphous carbon $f_{\rm aC}=0.3$,
and vacuum with $f_{\rm vac}=0.2$ to represent porosity, while ices
are ignored.

\subsection{Porosity \label{nk.sec}}

Optical constants $m= n - {\rm i} \ k$ with $k \ga 0$ of various grain
materials are displayed in Fig.~\ref{compnk.fig}. Optical constants
are shown for the aC-H$_2$ mixture of amorphous carbon by
\citet{Zubko96}, silicate and graphite by \citet{Draine03}, and a 97:3
mix in mass of amorphous silicate grains with 2\% vaccum inclusion
that are composed of MgO$-$0.5 SiO$_2$ and Mg$_{0.8}$Fe$^{2+}_{0.2}$
SiO$_3$ using optical constants by \citet{Demyk22} and as discussed in
Sect.~\ref{lab.sec}. The Si and aC grains differ by large in $n$ and
$k$, respectively.

By considering fluffy particles the optical constants are shifted to
lower values and follow a similar spectral shape as for pure materials
that do not have vacuum inclusions. This is seen in
Fig.~\ref{compnk.fig} by comparing aC-H$_2$ grains by \cite{Zubko96}
with fluffy aC \citep{KS94} with 20\% vaccum inclusion and using
Eq.~(\ref{Brugg.eq}). The spectral shape of the agglomerates by
\cite{DH21} is shown in Fig.~\ref{compnk.fig} in orange without (full
line) and with 20\% porosity (dashed), which can be compared with
fluffy composites in black. The latter is a representation of dark
dust. Again in the latter two sets similarities in the spectral shape
of $n$ and $k$ are visible as well as a larger shift in the FIR/submm
where differences in the slope become more pronounced. Also important
is that by changing the vacuum content of materials the spectral slope
remains unaltered. Thus the slope of the dust emissivity of a
material, e.g. in the submm, does not change with porosity. For the
same mass the strength of the dust emissivity increases by increasing
the porosity because fluffy grains are larger than particles of the
same material that have less or no vacuum inclusions.

\subsection{Aligned spheroidal grains \label{spheroids.sec}}

The interstellar polarization phenomenon in the optical and IR/mm
cannot be explained by spherical dust particles. Interstellar
polarization of starlight and polarized dust emission is explained by
partially aligned non-spherical dust grains that wobble and rotate
about the axis of the greatest moment of inertia. Most simple
representations of finite-sized non-spherical grains are
spheroids. They are characterised by the ratio $a/b$ between major and
minor semi-axes. Spheroids can be either in form of oblates such as
pancakes or prolates such as needles. The volume of a prolate is the
same as a sphere with a radius given by $r^3 = a \cdot b^2$.

The dust optical depth at frequency $\nu$ is given by the product of
the column density $N_{\rm d}$ of the dust along the sightline and the
dust extinction cross-section $C_{\rm ext} = C_{\rm abs} + C_{\rm
sca}$, which is the sum of absorption and scattering

\begin {equation}
\tau ({\nu}) = N_{\rm d} \ C_{\rm ext}({\nu}) \ . 
\label{tauCext.eq}
\end {equation}

\noindent
Linear polarization is

\begin {equation}
p(\nu) = N_{\rm d} \ C_{\rm p}({\nu}) \ , 
\label{pCp.eq}
\end {equation}

\noindent
where $C_{\rm {p}}$ is the linear polarization cross-section.

Physical processes that led to grain alignment are under debate and
various mechanisms have been proposed, see reviewed by \citet{V12} and
\citet{Andersson15}. In picket fence \citep{DB74} a fraction of the
grains is in perfect alignment while the other particles are randomly
oriented. A modified version of it is used by \citet{DH21}, where it
is assumed that the cross-sections along the ${\bf E}$ and ${\bf B}$
field depends on $\cos^2{\theta}$ of the scattering angle (the angle
between the wave vector ${\bf k}$ of the incoming light and the
symmetry axis of the grain). There exist also non-magnetic alignment
processes such as in gas streams of stellar winds or by an-isotropic
illumination where photons are predominantly absorbed from one side
of the particle. Radiative torque alignment has also become
popular. In the Milky Way the direction of the dust-induced
polarization in the optical is well correlated with the orientation
of the magnetic field, which is derived from synchrotron emission in
the radio. Hence magnetic fields are often taken to be responsible
for the grain alignment in the ISM. In that picture, questions arise
when estimating the relaxation times for the magnetic alignment of
the grain in comparison to the time when the disorder is again
established through collisions with gas atoms. For solving that
puzzle supra-thermal grain rotation (at a frequency of $10^5$\,Hz)
were proposed as well as (super-)paramagnetic or ferromagnetic
relaxation. In the following the imperfect Davies-Greenstein
mechanism (IDG) is applied in which Fe atoms in the dust interact
with the weak ${\bf B}$ field of the ISM. One advantage of IDG is
that the orientation of a spinning and wobbling spheroid can be
described mathematically in closed form. For angles $\psi$,
$\varphi$, precision angle $\beta$, and magnetic field direction
$\Omega$, applied as in the notation by \citet{HG80} and suffix TM
for the transverse magnetic and TE for transverse electric
polarization direction as in the notation by \citet{BH83}, the
alignment function becomes

\begin{equation}
{f}(\beta) = \frac{\frac{r + 0.1 \ \delta_0\ }{r +\delta_0} \, \sin \beta}{\left ( \left(\frac{r + 0.1 \ \delta_0\ }{r +\delta_0}\right)^2 \cos^2 \beta + \sin^2 \beta \right )^{3/2} } \ .
\end{equation}

\noindent The alignment efficiency $\delta_0 \sim 0.1 - 10\, \mu$m
\citep{Das10, S14} is related to the physical picture. It impacts the
maximum of the polarization $p_{\rm max}$ but not the shape
$p(\lambda)$ of the polarization (Voshchinnikov \& Das 2008). The
cross-sections are computed for given efficiency factors $Q$
(Sec.~\ref{cross.sec}) as average over the orientation of the
wobbling particles

\begin{equation}
{C}_{{\rm ext}}(\nu) = \frac{2}{\pi} 
\int (Q_{{\rm ext}}^{\rm{TM}} + Q_{{\rm ext}}^{\rm{TE}}) \, r^2 \,
f(\beta) \, \mathrm{d}{\varphi} \, \mathrm{d}{\Omega} \, \mathrm{d}{\beta} \ ,
\label{Cext.eq}
\end{equation}

\begin{equation}
{C}_{{\rm p}}(\nu) = \frac{1}{\pi} \int
(Q_{{\rm ext}}^{\rm{TM}} - Q_{{\rm ext}}^{\rm{TE}}) \, r^2 \,
f( \beta) \, \cos (2{\psi}) \, 
\mathrm{d}{\varphi} \, \mathrm{d}{\Omega} \, \mathrm{d}{\beta} \ .
\label{Cp.eq}
\end{equation}

\subsection{Reddening}

The mass extinction cross-sections $K_{{\rm {ext}}, i}(r)$\,
(cm$^2$/g-dust) of a dust particle of population $ i \in \{ {\rm {Si,
aC, vSi, VSG, Dark}}\}$, radius $r$, and density $\rho_i$\, is

\begin{equation}
K_{{\rm ext}, i}(r) = \frac {m_i} {{\displaystyle {\frac {4 \pi}{3}}
\ \rho_i}} \ \frac {r^{-q}} { \displaystyle
\int_{r_{-,i}}^{r_{+,i}} r^{3-q} \ \mathrm{d}r} \ C_{{\rm ext},
i}(r) \ .
\label{K.eq}
\end{equation}

\noindent
The cross-sections $C$ are derived using efficiency factors $Q$, which
are computed for spheres by Mie theory and spheroids as in
Sec.~\ref{spheroids.sec}. The relative weight, also called specific
mass, in 1\,g of dust in component $i$, is

\begin{equation}
m_{\rm {i}} = {\frac {\frac{[\rm{X}]}{[\rm{H}]}_{\rm {i}} \ \mu_{\rm X}}
{\left(\frac{[\rm{C}]}{[\rm{H}]}_{\rm {aC}} + \frac{[\rm{C}]}{[\rm{H}]}_{\rm {gr}} +
\frac{[\rm{C}]}{[\rm{H}]}_{\rm {PAH}}\right) \ \mu_{\rm C} +
\left({\frac{[\rm{Si}]}{[\rm{H}]}_{\rm {Si}}} + \frac{[\rm{Si}]}{[\rm{H}]}_{\rm {sSi}}\right)
\ \mu_{\rm Si}}} \ ,
\label{w.eq}
\end{equation}

\noindent
where relative dust abundances of element X, which is either C or Si,
with respect to H are denoted by $[\rm{X}]/[\rm{H}]_{\rm i}$ together
with a subscript $i$ for each of the dust population, $\mu_{\rm C}=12$
is the molecular weight of carbon, $\mu_{\rm Si}=134.46$ that of
silicate grains with bulk densities $\rho_C \sim 1.6$\,g/cm$^3$ and
$\rho_{\rm Si} \sim 3.41$\,g/cm$^3$ \citep{DH21}, respectively. The
specific mass for PAH is computed in similar terms
\citep{S14}. Specifying the mass extinction cross-section per gram of
dust has the advantage that only relative element abundances
$[\rm{X}]/[\rm{H}]_{\rm i}$ need to be specified instead of the more
uncertain absolute solid-phase element abundances (in ppm), which
however, may be used as a guideline. The total mass extinction
cross-section averaged over the dust size distribution in
cm$^2$/g-dust is

\begin{equation}
K_{\rm ext} = \sum_i \ \int_{r_-}^{r_+} K_{{\rm ext}, i}(r)
\ \mathrm{d}r \ .
\label{Kext.eq}
\end{equation}

\noindent
The wavelength dependence in the above expressions (Eq.~\ref{Cext.eq}
- \ref{Kext.eq}) is dropped for clarity. The reddening curve of the
dust model is derived using Eqs.~(\ref{Rv.eq}, \ref{tau.eq}) and
noting $K$ as the mass extinction cross-section

\begin{equation} \label{Emod.eq}
E(\lambda) = \frac {K_{\rm{V}} } {K_{\rm {B}} - K_{\rm
{V}}} \left( \frac {K(\lambda)} {K_{\rm {V}}} - 1 \right) \ .
\end{equation}

\noindent
Note by using $R_{\rm V}$ (Eq.~\ref{Rv.eq}) that the reddening
$E(\lambda)$ is derived self-consistently from the dust model without
extrapolation to infinite wavelengths.

\subsection{Emission}

The emission $\epsilon_{\nu , i} (r)$ of a dust particle of radius
$r$, grain material $i$ at frequency $\nu$ is

\begin{equation}
\epsilon_{\nu , i}(r) = K_{\nu , i}^{abs} (r) \ \mathlarger{\int}
{P_i(r,T) \, B_{\nu}(T) \, \mathrm{d}T } \,,
\label{emis.eq}
\end{equation}

\noindent where $K_{\nu , i}^{abs}$ is the mass absorption
cross-section (Eq.~\ref{K.eq}), $B_{\nu}(T)$ is the Planck function,
and $P_i(r,T)$ is the temperature distribution function that gives the
probability of finding a particle of material $i$ and radius $r$ at
temperature $T$. This function is determined from the energy balance
between the emission and absorption of photons from the mean intensity
$J_{\nu}$, for which here the ISRF by \cite{Mathis83} is applied

\begin{equation}
\mathlarger{\int} {K_{\nu , i}^{abs} (r) \, J_{\nu}(T) \,
\mathrm{d}\nu } = \mathlarger{\int} {K_{\nu , i}^{abs} (r)
\, P_i(r,T) \, B_{\nu}(T) \ \mathrm{d}T \, \mathrm{d}\nu } \ .
\end{equation}

\noindent It is evaluated using an iterative scheme that is described
by \citet{K08}. The $P(T)$ function only needs to be evaluated for
small grains as it approaches a $\delta$-function for large particles
where the temperature fluctuates very little around the equilibrium
temperature. The total emission $\epsilon_{\nu}$ of the dust at
frequency $\nu$ is given as the sum of the emission $\epsilon_{\nu,
i}(r)$ of all dust components.

\subsection {Gas-to-dust mass ratio}

Observations of the diffuse emission of the galaxy are given per
hydrogen column density, $\lambda \ I_{\lambda} / N_{\rm {H}}$
(erg/s/sr/\,H-atom), whereas the dust emission of the model is
computed (Eq.~(\ref{emis.eq}) per dust mass (erg/s/sr/\,g-dust). For
the necessary conversion of the dust to the gas column densities often
a procedure is applied assuming that for sightlines in the diffuse ISM
the extinction is proportional to the hydrogen column density $A_{\rm
V} = 1.086 \times \tau_{\rm V} \propto N_{\rm H}$. In addition, one
presumes that $A_{\rm V}$ can be reasonably estimated by applying a mean
value of the total-to-selective extinction and the derived reddening
using $A_{\rm V} = {R_{\rm V}} \times {E(\rm{B} - \rm{V})}$. Further,
one assumes that the ratio $\zeta = N_{\rm H}/ {E(\rm{B} - \rm{V})}$
stays roughly constant in the ISM. Traditionally $\zeta= 5.8$
($10^{21}\, \rm{H} \, \rm{cm}^{-2} \, \rm{mag}^{-1}$) is applied,
which is derived from 75 sightlines observed within 3.4\,kpc by
Copernicus \citep{Bohlin78}. For translucent clouds at $\tau_{\rm V}
\ga 0.5$ observed by FUSE $\zeta = 5.94 \pm 0.37$, from here onwards
in units as before \citep{Rachford09}, and from X-ray observations
$\zeta \sim 6.3-6.5$ \citep{Zhu17} using ${R_{\rm V}}=3.1$. However,
striking differences are derived from radio observations of HI with
lower values in earlier work of $\zeta = 4.6$ \citep{Mirabel79} and
5.1 \citep{Knapp74} or higher values in recent studies of $\zeta =
8.3$ \citep{Liszt14}, 8.8 \citep{Lenz17} and 9.4
\citep{Nguyen18}. Dust emission of the diffuse ISM is observed towards
high galactic latitudes where there is less dust per H atom as in the Galactic plane so that the latter value of $\zeta$ has been favoured
by \citet{HD21}.

In the procedure applied here these uncertainties in $\zeta$, $R_{\rm
V}$, and ${E(\rm{B} - \rm{V})}$ are avoided. A gas-to-dust mass
ratio ${M_{\rm {gas}}}/{M_{\rm {dust}}}$ is introduced that is given
by scaling the dust emission spectrum of Eq.~(\ref{emis.eq}) to the
Planck data at 350\, $\mu$m. The observed emission is fit by

\begin{equation}
\frac{\lambda \, I_{\lambda}}{N_{\rm {H}}} = \frac{M_{\rm
{dust}}}{M_{\rm {gas}}} \ \frac{\lambda
\ \epsilon_{\lambda}}{m_{\rm p}} \ , 
\label{mgd.eq}
\end{equation}

\noindent
with H-atom mass $m_{\rm p}$. The procedure has the advantage that
${M_{\rm {gas}}}/{M_{\rm {dust}}}$ is derived in the same direction as
to where the data were taken. Of course, uncertainties that arise due
to possible variations in the strength of the ISRF towards the
observed fields remain and are assumed to be small.

\subsection{Polarized extinction}

Starlight polarization is derived by dichroic extinction of the dust
(Eqs.~\ref{tauCext.eq} - \ref{pCp.eq}, \ref{Cp.eq})

\begin{equation}
p(\nu)/\tau_{\rm V} = K_{\rm {p}}(\nu) / K_{\rm {ext, V}} \ ,
\label{pmod.eq}
\end{equation}

\noindent
$K_{\rm p}$ denotes the total linear mass polarization cross-section
(cm$^2$/g-dust) and is computed similar to $K_{\rm ext}$ by replacing
in Eq.~(\ref{K.eq}) the cross-section $C_{{\rm ext}, i}$
(Eq.~\ref{Cext.eq}) with $C_{{\rm p}, i}$ (Eq.~\ref{Cp.eq}).

\subsection{Polarized emission}

Polarized dust emission of component $i$ is computed by integrating over the
minimum ${r_{pol , i}^-}$ to maximum ${r_{pol , i}^+}$ alignment radii

\begin{equation}
\epsilon_{pol, i} {(\nu)} = \mathlarger{\int}_{r_{pol ,
i}^{-}}^{r_{pol , i}^+} K_{pol, i} (\nu , r) \, B_{\nu}(T) \, \mathrm{d}r \ .
\label{emispol.eq}
\end{equation}

\noindent
Note that $C_{\rm p} = K_{\rm p} = 0$ for spherical or non-aligned
grains. The total polarized dust emission $\epsilon_{\rm {pol}}$ is
given as the sum of the polarized emission $\epsilon_{pol, i}$ of all
components contributing to the polarization. These are large aC and Si
prolate grains and dark dust when also considered of prolate shape. Polarization by nanoparticles is not considered.

\section{Method}

The dust model is confronted with the observational constraints
presented in Sec.~\ref{obs.sec}. As described by \cite{Zubko04} the
fitting procedure leads to a typical ill-posed inversion problem where
the solution is extremely sensitive to small changes in input data and
with several priors such as the size distribution or grain composition
as unknowns. Therefore, a least $\chi^2$-technique is applied for
which first, a method for finding the best fit of the reddening curve
is presented that respects dust abundance constraints
(Eq.~\ref{abu.eq}). For that model, the parameters which impact the
shape of the starlight polarization are varied to find the best fit to
the mean Serkowski curve. The latter model is then compared to the
diffuse galactic emission, which requires applying a gas-to-dust mass
ratio. The flatness in the submm/mm polarization spectrum as well as
the ratio of the starlight polarization to submm polarization is
tested. The parameter space of the model is explored and the goodness
of the fit is quantified as the sum of differences between
observations and the model, each squared and divided by the observed
data. For the reddening curve the goodness parameter is denoted by
$\chi^2_{\rm r}$ and for the polarized extinction $\chi^2_{\rm p}$,
respectively.

\subsection{Cross-sections \label{cross.sec}}

The various dust cross-sections for extinction, scattering, and
polarization are computed in the spectral range between $90\,\rm{nm}
\la \lambda \la 1$\,cm and for grain radii range between {5\,\AA \, $
  \la r \la 10\,\mu$m}. The challenge is computing the efficiency
factors $Q = C/\pi r^{2}$ for the two polarization directions of
elongated particles (Eqs.~\ref{Cext.eq},~\ref{Cp.eq}). For small size
parameter $x =2 \pi r /\lambda <<1$ the Rayleigh approximation might
be used \citep{K08}. For typical ISM grains with sizes $r \leq
0.3\,\mu$m scattering becomes small at $\lambda \simgreat 8\, \mu$m
and the Rayleigh limit $2 \pi r << \lambda$ is held. The Rayleigh
approximation breaks at $x \simgreat 0.25$ \citep{V04,DH21shape} and
cannot be used in most of the interesting cases. Electromagnetic
absorption and scattering by spheroids can be treated using different
methods. A numerical solution of separation of variables in the
Maxwell equation is presented by \citet{VF93}. The discrete dipole
approximation by \citet{DF94} offers a possibility for the treatment
of various grain structures. In addition, the extended boundary
condition method by \citet{M00} also known as T-matrix approximation
is available. The methods have been tested extensively by the authors
and excellent agreement at $x \sim 10$ is exemplified
\citep{DH21shape}. The program{\footnote {Fortran code SPVV8.5 kindly
  provided by N. Voshchinnikov$^{\dagger}$.}} by \citet{VF93} is
used. It returns for a given complex optical constant $m$ and the size
parameter $x$, the efficiency factors $Q_{{\rm ext}}^{\rm{TM}}(x,m)$
and $Q_{{\rm ext}}^{\rm{TE}}(x,m)$ (Eqs.~\ref{Cext.eq} -
\ref{Cp.eq}). The code converges up to $\lvert m - 1 \rvert \ x \ \la
25$, which is found to be at larger $x$ than in most of the other
applications.

\subsection{Fitting procedure \label{fitting.sec}}

The radial grid in grain size is ascending using $r_{i+1} = 1.05
\ r_i$. Intermediate radii between two consecutive grid points are
linearly interpolated when necessary. The upper radius of the dark dust
is varied for six values of $r_{\rm {Dark}}^{+} = 0, 0.5, 0.8, 1.0,
5.0$, and $10\,\mu$m, where $r_{\rm {Dark}}^{+} = 0\,\mu$m refers to a
model without dark dust. Models assuming even larger grains were also
inspected without finding striking features. By assuming a continuous
distribution in the particle size the minimum radius of the dark dust
follows that of the maximum radius of the large aC and Si grains,
hence $r_{\rm {Dark}}^{-} = \rm{max}(r_{\rm {ac}}^{+},r_{\rm
{Si}}^{+})$. Models fitting the observational constraints show
$r_{\rm {ac}}^{+} < r_{\rm {Si}}^{+}$. The exponent of the size
distribution $q$ is kept as a free parameter, however, the same $q$ is
used for all three dust populations. After some experiments the mass
in dark dust is adopted to be 10\,\% of the total dust mass; this
parameter will be fine-tuned when fitting the dust emission in the mm
range. Dust abundances are given in the model in relative terms and
one parameter can be kept fixed, and $[\rm{Si}]/[\rm{H}]_{\rm {Si}}=
15$\,ppm is set.

The polarization data are solely fit by large grains unless otherwise
specified. A detailed fit to the Serkowski curve is obtained when the
alignment of both large grain materials aC and Si differ. Models
fitting the absolute values of the polarization shall be applied
towards individual sightlines that are dominated by a single absorbing
cloud \citep{S18}. Hence parameters that change $p_{\rm{max}}$ but do
not affect the spectral shape of the polarization are held constant
\citep{V12}. This concerns the magnetic field direction that is set to
$\Omega =60^{\rm{o}}$, the ratio of the major axis of prolates $a/b
=2$, and the alignment efficiency of aC grains $\delta_0 \, (\rm
{aC})= 10\,\mu$m, while variations of Si grains of $\delta_0 \, (\rm
{Si}) = 0.5, 1.0$, and $10\,\mu$m are explored. The maximum alignment
radius $r_{\rm {pol}}^{+}$ shows no strong variations of the starlight
polarization spectrum whereas the minimum alignment radius $r_{\rm
{pol}}^{+}$ has a stong impact \citep{S14}. As maximum alignment
radius the upper radius derived from the reddening fit is applied,
$r_{\rm {pol, Si}}^{+} = r_{\rm {Si}}^{+}$ and $r_{\rm {pol, aC}}^{+}
= r_{\rm {aC}}^{+}$, respectively. The minimum alignment radius is
taken as a free parameter for aC and Si grains. The Si grains are
dominating the polarization curve at shorter wavelengths and generally
$r_{\rm {pol, Si}}^{-} \la r_{\rm {pol, aC}}^{-}$ is hold.

The reddening curve is fit using the $3 \times 6 $ values as above for
$\delta_0$ and $r_{\rm {Dark}}^{+}$, which have minor impact on
$E_{\rm{mod}}(\lambda)$ and in addition the seven parameters of the
dust model: $[\rm{C}]/[\rm{H}]_{\rm {aC}}$, $[\rm{C}]/[\rm{H}]_{\rm
{VSG}}$, $[\rm{C}]/[\rm{H}]_{\rm {PAH}}$, $[\rm{Si}]/[\rm{H}]_{\rm
{vSi}}$, $q$, $r_{\rm {ac}}^{+}$, and $r_{\rm {Si}}^{+}$. A set of
best fit parameters are computed by a least $\chi^2$-technique
utilizing the Levenberg--Marquardt algorithm as implemented in
MPFIT\footnote{http://purl.com/net/mpfit} \citep{Markwardt09}. The algorithm can find local minima. The challenge is to identify
the global $\chi^2_{\rm r}(r)$ minimum of the reddening curve
fit. This minimum is derived by starting the algorithm using many
different initial parameter values. As an initial guess the mean dust
parameters of ISM sightlines fit by \cite{S18} are applied:
$[\rm{C}]/[\rm{H}]_{\rm {aC}} =67$, $[\rm{Si}]/[\rm{H}]_{\rm
{vSi}}=5$, $[\rm{C}]/[\rm{H}]_{\rm {VSG}}=17$,
$[\rm{C}]/[\rm{H}]_{\rm {PAH}}=10$\, (ppm), and $q=3$. In addition one
initial radius $r_{\rm i}$ is selected out of 12 different upper sizes
of large aC and Si grains of the radial grid between $ 180 \la r_{\rm
i} = r_{\rm {ac}}^{+} = r_{\rm {Si}}^{+} \la 300$\,(nm). This limit
is based on an exploration of the model space which has shown that
start values of $r_{\rm i} > 300$\,nm did not result in acceptable
solutions. An initial MPFIT run is started keeping $r_{\rm {ac}}^{+} =
r_{\rm {Si}}^{+}$ fixed and after converging the resulting 5
parameters $[\rm{C}]/[\rm{H}]_{\rm {aC}}$, $[\rm{C}]/[\rm{H}]_{\rm
{VSG}}$, $[\rm{C}]/[\rm{H}]_{\rm {PAH}}$, $[\rm{C}]/[\rm{H}]_{\rm
{vSi}}$, and $q$ are held constant, while MPFIT is re-started a
second time with $r_{\rm {ac}}^{+}$ and $r_{\rm {Si}}^{+}$ set as
free.
%%%%%%%%%%%%%%%%%%%%%%%%%%%%%%
\begin{table*}[!htb]
\scriptsize
\begin{center}
\caption {Parameters of the dark dust model for the diffuse ISM using
\citet{Draine03} optical constants for amorphous silicate
grains. \label{para.tab}}
\begin{tabular}{c c | c c r c r r | c c c c c c| c c c c}
\hline\hline
1 & 2 & 3 & 4 & 5 & 6 & 7 & 8 & 9 & 10 & 11 & 12 & 13 & 14 & 15 & 16 & 17 & 18 \\
\hline
& & & & & & & & & & & & & & & & & \\
\multicolumn{2}{c|}{Model} & \multicolumn{6}{c|}{Mass ratio (\%)} & \multicolumn{6}{c|}{Sizes } & \multicolumn{4}{c}{Derived quantities} \\
& & \multicolumn{6}{c|}{Abundances (ppm)} & & & & & & & & & & \\
\hline
$r_{\rm {Dark}}^{+}$ &
$\chi^2$ &
$m_{\rm {Dark}}$ &
$m_{\rm {Si}} $ &
$m_{\rm {vSi}} $ &
$m_{\rm {aC}} $ &
$m_{\rm {VSG}} $ &
$m_{\rm {PAH}} $ &
$q$ &
$r_{\rm {Si}}^{+}$ &
$r_{\rm {ac}}^{+}$ &
$r_{\rm {pol, Si}}^{-}$ &
$r_{\rm {pol, aC}}^{-}$ &
$\delta_0 (\rm{Si})$ &
$\frac{[\rm{C}]}{[\rm{Si}]}$ &
$\frac{M_{\rm {gas}}}{M_{\rm {dust}}}$ &
$R_{\rm {V}}$ &
$\tau_{\rm {V}}$ \\
% ------------------ 3 line
($\mu$m)& $N \ (\chi^2<1)$ & &
$\frac{[\rm{Si}]}{[\rm{H}]}_{\rm {Si}}$ & $\frac{[\rm{Si}]}{[\rm{H}]}_{\rm {vSi}}$&
$\frac{[\rm{C}]}{[\rm{H}]}_{\rm {aC}}$ & $\frac{[\rm{C}]}{[\rm{H}]}_{\rm {VSG}}$ &
$\frac{[\rm{C}]}{[\rm{H}]}_{\rm {PAH}}$ & & (nm) & (nm) & (nm) & (nm) & ($\mu$m) & & & \\
\hline
& & & & & & & & & & & & & & & & & \\
0$^{\ddagger}$ & 0.5 & - & 46 & 23 & 25 & 4 & 3 & 3.0 & 243 & 203 & 80& 166 & 0.5 & 5.2 & 125 & 3.09 & 1.01 \\
& 0 & & 15 & 7 & 92 & 15 & 10 & & & & & & & & & & \\
& & & & & & & & & & & & & & & & & \\
0.75$^{\dagger}$& 0.7 & $7$ & 43 & 20 & 24 & 4 & 2 & 3.0 & 228 & 201 & 84& 166 & 0.5 & {\it{5.5}} & 138 & 3.16 & 1.42 \\
& 0 & & 15 & 7 & 96 & 15 & 8 & & & & & & & & & & \\
\hline
& & & & & & & & & & & & & & & & & \\
1 & 0.6 & $6$ & 44 & 21 & 23 & 4 & 2 & 3.0 & 217 & 207 & 76& 143 & 1.0 & 5.0 & 131 & 3.09 & 1.37 \\
& 5 & & 15 & 7 & 88 & 15 & 8 & & & & & & & & & & \\
& & & & & & & & & & & & & & & & & \\ 
5 & 0.6 & $7$ & 41 & 22 & 23 & 3 & 3 & 2.9 & 220 & 208 & 80& 166 & 0.5 & 5.2 & 124 & 3.11 & 1.30 \\
& 3 & & 15 & 8 & 95 & 13 &11 & & & & & & & & & & \\
& & & & & & & & & & & & & & & & & \\ 
10 & 0.7 & $6$ & 49 & 17 & 22 & 3 & 2 & 3.1 & 239 & 222 & 88& 183 & 0.5 & 4.6 & 124 & 3.18 & 1.36 \\
& 3 & & 15 & 6 & 91 & 14 & 8 & & & & & & & & & & \\
\hline
\end{tabular}
\end{center}
{{\bf Notes:} In Col.~(1) the upper radius of the dark dust
  agglomerates $r_{\rm {Dark}}^{+}$ is specified, in Col.~(2) the
  goodness $\chi^2$ of the best fit model and below the number $N$ of
  models that are consistent with the observational constraints and
  fit the reddening and Serkowski curve at $\chi^2 <1$. Corresponding
  parameters of the specific mass in \% per gram dust (Col.~3-8) of
  dark dust as lower limit $m_{\rm {Dark}}$, and for large silicates
  $m_{\rm {Si}} $, nano-sized silicates $m_{\rm {vSi}}$, large
  amorphous carbon $m_{\rm {aC}} $, very small graphite $m_{\rm {VSG}}
  $, and PAHs $m_{\rm {PAH}}$. Below, in the second row of Col.~3-8,
  exemplified dust abundances [X]/[H] (ppm) are given by adopting
  [Si]/[H] = 15\,ppm in large silicates. Col.~(9) gives the exponent
  of the dust size distribution $q$. The upper radius of large
  silicates $r_{\rm {Si}}^{+}$ (Col.~10) and amorphous carbon $r_{\rm
    {ac}}^{+}$ (Col.~11), their lower alignment radii $r_{\rm {pol,
      Si}}^{-}$ and $r_{\rm {pol, aC}}^{-}$ (Col.~12, 13),
  respectively, and the alignment efficiency (Col.~14) of large
  silicates $\delta_0 (\rm{Si})$ is given. Derived quantities of the
  dust models are given for the dust abundance ratio [C]/[Si]
  (Col.~15), the gas-to-dust mass ratio $M_{\rm {gas}}/M_{\rm {dust}}$
  (Col.~16), the total-to-selective extinction $R_{\rm {V}}$
  (Eq.~\ref{Rv.eq}, Col.~17), and the optical depth (Col.~18) that
  matches the optical-to-submm polarization ratio of ${p_{850\,\mu{\rm
        m}}} / {(p_{\rm {V}}/\tau_{\rm {V}})} = 4.31$
  (Eq.~\ref{Pratio.eq}) by \citet{Planck20}.
  \newline $^{\dagger}$
  Model violates abundance constrain (Eq.~\ref{abu.eq}).
  \newline
  $^{\ddagger}$ Models without dark dust do not fit
  the \cite{Planck15, Planck20} data at $\ga 0.8$\,mm.}
\end{table*}

%%%%%%%%%%%%%%%%%%%%%%%%%%%%%%%%%%%%%%%%%%%%%%%% 

This procedure is iterated, generally twice, until $\chi^2_{\rm r}$ is
not reduced further. Then the starlight polarization curve is fit
minimizing $\chi_{\rm p}^2 \left(\delta_0 \, (\rm {Si})\,, r_{\rm
{pol}}^{-}\,(\rm {Si})\,, r_{\rm {pol}}^{-}\,(\rm {aC}) \right)$
between the observed Serkowski curve (Eq.~\ref{Serk.eq}) and the dust
model (Eq.~\ref{pmod.eq}). In that method the 7 best-fit dust
parameters derived from the reddening curve procedure are held
constant and $\chi_{\rm p}^2$ is computed for the three $\delta_0 \,
(\rm {Si})$ and all combinations of the minimum alignment radii of aC
and Si grains between {$6\,{\rm{nm}} < r_{\rm {pol}}^{-} <
\rm{min}(r^{+}_{\rm {aC,Si}})$}, respectively.

For keeping the computational time within reasonable limits
the fitting procedure is vectorized by running calls to MPFIT with the
many different start values parallel.

%%%%%%%%%%%%%%%%%%%%%%%%%%%%%%%%%%%%%%%%%%%%%%%%%

\section{Results}

A $r_{\rm {Dark}}^{+} \times \delta_0 (\rm{Si}) \times r_{\rm
{ac}}^{+}$ tuple with $3 \times 6 \times 12$ models is computed
applying the procedure of Sec.~\ref{fitting.sec}. The model grid
includes for each combination of $\delta_0 (\rm{Si})$ and $r_{\rm
{Dark}}^{+}$, the seven derived best-fit parameters to the reddening
$[\rm{C}]/[\rm{H}]_{\rm {aC}}$, $[\rm{C}]/[\rm{H}]_{\rm {VSG}}$,
$[\rm{C}]/[\rm{H}]_{\rm {PAH}}$, $[\rm{C}]/[\rm{H}]_{\rm {vSi}}$, $q$,
$r_{\rm {ac}}^{+}$, $r_{\rm {Si}}^{+}$, the two best fit parameters to
the starlight polarization $r_{\rm {ac}}^{+}$, $r_{\rm {Si}}^{+}$, and
the corresponding goodness parameters $\chi^2_{\rm r}$ and
$\chi^2_{\rm p}$, respectively. The majority of models provide a
reasonable fit to the reddening and polarization curves but do not
respect the abundance constraint of $\rm{[C]/[Si]} \la 5.25$
(Eq.~\ref{abu.eq}) and are rejected. Out of the 216 models, 27 are
consistent with the dust abundance ratio (Eq.~\ref{abu.eq}) and 22
remain after applying a $3 \ \sigma$ outlier rejection in $\chi^2_{\rm
r}$ and $\chi^2_{\rm p}$, respectively. Their $\chi^2_{\rm r}$ and
$\chi^2_{\rm p}$ are each normalised to a mean of one so that they can
be combined. The model that simultaneously fits both curves best is
selected from a minimum $\chi^2$ condition where each parameter is
given the same weight using

\begin{equation}
\chi^2 (r_{\rm {Dark}}^{+}) = \frac{7}{10} \ \chi^2_{\rm r} \ + \ \frac{3}{10}
\ \chi^2_{\rm p} \ .
\end{equation}

\begin{figure}[htb!]
\includegraphics[width=9.9cm,clip=true,trim=4.8cm 8.cm 2.2cm 8.4cm]{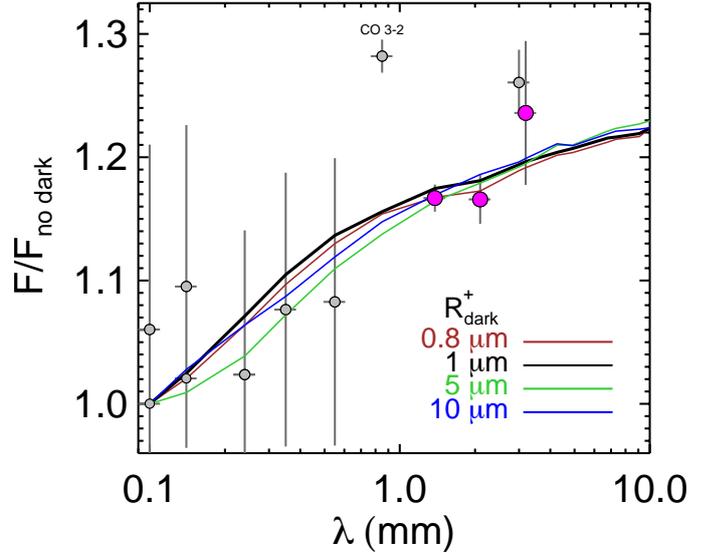}
\caption {Ratio $F / F_{{\rm {no}} \ {\rm {dark}}} $ of the dark dust
models and the observed fluxes (circles, using data as in
Fig.~\ref{sed.fig}) divided by the model neglecting dark dust
($r_{\rm {Dark}}^{+} =0\,\mu$m). Best fits (lines) to the 1.4\,mm
and 2.1\,mm data derived by varying the amount of dark dust $m_{\rm
{Dark}}$ for models with $r_{\rm {Dark}}^{+} \simgreat 0.75\,\mu$m
as labelled. Parameters as in
Table~\ref{para.tab}. \label{xFIR.fig}}
\end{figure}

Model parameters with the minimum in $\chi^2 < 1$ for given $r_{\rm
{Dark}}^{+}$ are listed in Table~\ref{para.tab}. These models agree
with the observed reddening and polarization. Models that ignore dark
dust ($r_{\rm {Dark}}^{+}=0\,\mu$m) underpredict the observed emission
in the Planck bands at wavelength $\ga 0.8$\,mm
(Fig.~\ref{xFIR.fig}). There is no model with $r_{\rm {Dark}}^{+} =
0.5 \,\mu$m that fits reddening and polarization at $\chi^2 \la 1$.

%%%%%%%%%%%%%%%%%%%%%%%%%%%%%%%%%%%%%%%%%%%%%%%%%%
%%%%%%%%%%%%%%%%%%%%%%%%%%%%%%%%%%%%%%%%%%%%%%%%%%

For $r_{\rm {Dark}}^{+} = 0.75 \,\mu$m there is only one such model,
it has $\chi^2 = 0.7 $ but $\rm{[C]/[Si]} = 5.5$
(Table~\ref{para.tab}, Col.~15). Hence it violates the abundance
constraint (Eq.~\ref{abu.eq}). Therefore,  at $r_{\rm {Dark}}^{+} < 1
\,\mu$m no model is consistent with the observations at
$\chi^2<1$ and the number of such models is $N \ (\chi^2<1) =0$
(Table~\ref{para.tab}, Col.~2).

Models that agree with the observing constraints consider
micrometre-sized dark dust particles at $r_{\rm {Dark}}^{+} \ga 1
\,\mu$m. The global minimum over all models is found for $r_{\rm
  {Dark}}^{+} =1\,\mu$m at $\chi^2 = 0.6$. For this radius, there are
5 models at $\chi^2 < 1$. The best fitting models of $r_{\rm
  {Dark}}^{+} =5$ and $10\,\mu$m have similar $\chi^2 \ $ ($ \la
0.7$), respectively. Their fits to the observed reddening,
polarization, and dust emission are of similar quality as for the best
fit $r_{\rm {Dark}}^{+} =1\,\mu$m model. For each of these radii there
are $N \ (\chi^2<1) = 3$ models that fit the reddening and
polarization at $\chi^2<1$ and fulfill the abundance constraint
[C]/[Si] < 5.25 (Eq.~\ref{abu.eq}). The scatter in the individual
parameters between the selected models is small
(Table~\ref{para.tab}). The peak-to-peak variations in the abundances
for a given dust population (Table~\ref{para.tab}, Col.~3-8) and the
particular size parameters stay well below 10\,\%
(Table~\ref{para.tab}, Col.~10-13).

The fit of the $r_{\rm {Dark}}^{+} = 1\,\mu$m model with parameters of
Table~\ref{para.tab} to the observed reddening and extinction is shown
in Fig.~\ref{redd.fig}. The individual dust components show the
known behaviour of the reddening: nanoparticles are responsible for
the rise in the far UV, PAH and VSG fit the 2175\AA\ extinction bump,
and large aC and Si grains give a rather flat contribution from the U
band to shorter wavelengths. Remarkable is the constant reddening of
dark dust from the far UV to the NIR, up to wavelengths, which are
similar to the upper grain radius $r_{\rm {Dark}}^{+}$, which is for
that model at $\sim 1\,\mu$m (Fig.~\ref{redd.fig}, right). The
non-dispersed reddening provided by dark dust to the UBVRI bands is
significant. Dark dust causes extra dimming of light not accounted for
in dust models omitting this component.

The model fit to the observed starlight polarization curve is shown in
Fig.~\ref{Serk.fig}. For deriving such a detailed fit it is
necessary that both large Si and aC grains contribute to the
polarization and that both dust materials have distinct
characteristics in the alignment radii and the polarization efficiency
(cmp. Col~10-14 of Table~\ref{para.tab}). Polarized extinction is
dominated in the UV by Si grains and in the R band by aC grains.

Dark dust treated as prolates with axial ratio $a/b=2$,
$\delta_o=10\,\mu$m, and $r_{\rm {Dark}}^{+} = 1\,\mu$m contribute to
$p/p_{\rm {max}}$ most near $\sim 2\,\mu$m at less than a few percent.

Dark dust polarized extinction in the MIR is shown for models with
$r_{\rm {Dark}}^{+} = 1\,\mu$m and $r_{\rm {Dark}}^{+} = 10\,\mu$m in
Fig.~\ref{mirpol.fig}, respectively. Large aC grains show a flat
polarization at $\sim 20$\,\% of the maximum MIR polarization $p_{\rm{
max}}$, while Si grains dominate the spectrum. Dark dust, even when
elongated and aligned particles are assumed, contributes to the
normalised MIR polarization by less than $p/p_{\rm{ max}} <
10$\,\%. The composite polarization spectrum derived from observations
of two Wolf-Rayet stars refers to greatly distinct environments from
the diffuse ISM. Nevertheless, the data are shown in
Fig.~\ref{mirpol.fig} demonstrates the capabilities of using MIR
polarization as a method to estimate the stoichiometry of silicate
grains.

\begin{figure}[htb!]
\includegraphics[width=9.cm,clip=true,trim=4.cm 8.cm 3.cm 8.4cm]{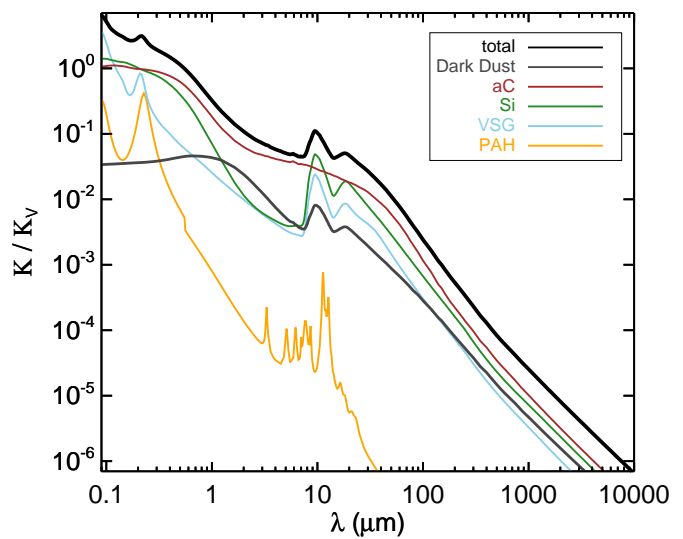}
\caption {Total extinction cross-section normalized to V for the
{$r_{\rm {Dark}}^{+} = 1 \, \mu$m} model (Table~\ref{para.tab}) with
individual components as labelled.\label{kappa.fig}}
\end{figure}

The dust emission of the models is compared to the diffuse emission of
the ISM that is observed at high galactic latitudes normalised per H
atom and is shown for the best fit $r_{\rm {Dark}}^{+} = 1\,\mu$m
model in Fig.~\ref{sed.fig}. The models are scaled to the 0.3\,mm
data, which allows for an estimate of the gas-to-mass ratio
(Eq.~\ref{mgd.eq}). A mean of ${M_{\rm {gas}}}/{M_{\rm {dust}}} = 126
\pm 4$ is derived for the best fitting models (Table~\ref{para.tab},
Col~16). The emission of the individual dust components shows the
known behaviour: nanoparticles are responsible for the MIR emission,
PAH to the IR emission bands and large grains dominate the FIR/submm.
The lowest temperature of aC grains is 16.8\,K, that of Si grains
14.8\,K, and dark dust is as cold as 12.3\,K for the $r_{\rm
{Dark}}^{+} = 1\,\mu$m model and 8.1\,K for the $r_{\rm {Dark}}^{+}
= 10\,\mu$m, respectively. Dark dust emission peaks at $\sim 0.2$\,mm
and becomes a more important contributor in the mm range where it even
overshines the emission by large Si grains. This is as expected by
comparing the extinction cross-sections of the different dust
ingredients. Noticeable in Fig.~\ref{kappa.fig} are the various PAH
features, the 9.7\,$\mu$m and 18\,$\mu$m bands of the silicates, and
that aC grains dominate the cross-sections over the entire spectrum at
$\lambda \ga 0.1\,\mu$m, except near the $9.7\,\mu$m
band. Furthermore, in the submm, the cross-section of dark dust has
similar strength as that of Si grains. Dark dust shows in the mm range
a slightly shallower decline in the emissivity than the other grains,
which is as expected inspecting the imaginary part $k$ of the optical
constants (Sec.~\ref{nk.sec}).

%%%%%%%%%%%%%%%%%%%%%%%%%%%%%%%%%%%%%%%%%%%%%%%%%%

%%%%%%%%%%%%%%%%%%%%%%%%%%%%%%%%%%%%%%%%%%%%%%%%%%

Models that exclude dark dust or models that do not alter the FIR/mm
spectral index of the grain emissivity do not account for the dust
emission observed by \citet{Planck15, Planck20}. This is exemplified
in Fig.~\ref{xFIR.fig}. The observed flux $F$ is divided by the
flux $F_{{\rm {no}} \ {\rm {dark}}}$ of the models that do not
consider dark dust ($r_{\rm {Dark}}^{+} = 0\,\mu$m,
Table~\ref{para.tab}). The latter is a kind of \cite{S14} model; they
include non-porous prolate grains and fit the DIRBE and Planck data up
to 0.35\,mm. These models are at $F/F_{{\rm {no}} \ {\rm {dark}}} \sim
1$ in Fig.~\ref{xFIR.fig}. They systematically underpredict the
emission in the Planck bands at $0.8 - 3$\,mm by $15 - 30$\,\%, which
is significant considering the unprecedented precision of the Planck
data. Fluffy and spheroidal grains show greater dust emissivity in the
FIR/mm range than do non-porous and spherical grains of the same
mass. The FIR/mm spectral slope of the dust emissivity stays
invariable by increasing the porosity or the axial ratio of the
spheroidal particles (cmp. Fig.~\ref{compnk.fig} and Fig.~3 in
\cite{S14}). Hence for fitting the emission in the mm range one must
either change the spectral index of the dust emissivity by a suitable
set of optical constants \citep{DH21} or include dark dust.

%%%%%%%%%%%%%%%%%%%%%%%%%%%%%%%%%%%%%%%%%%%%%%%%%%

\begin{figure}[htb]
\includegraphics[width=9.cm,clip=true,trim=4.cm 7.5cm 3.5cm
8.4cm]{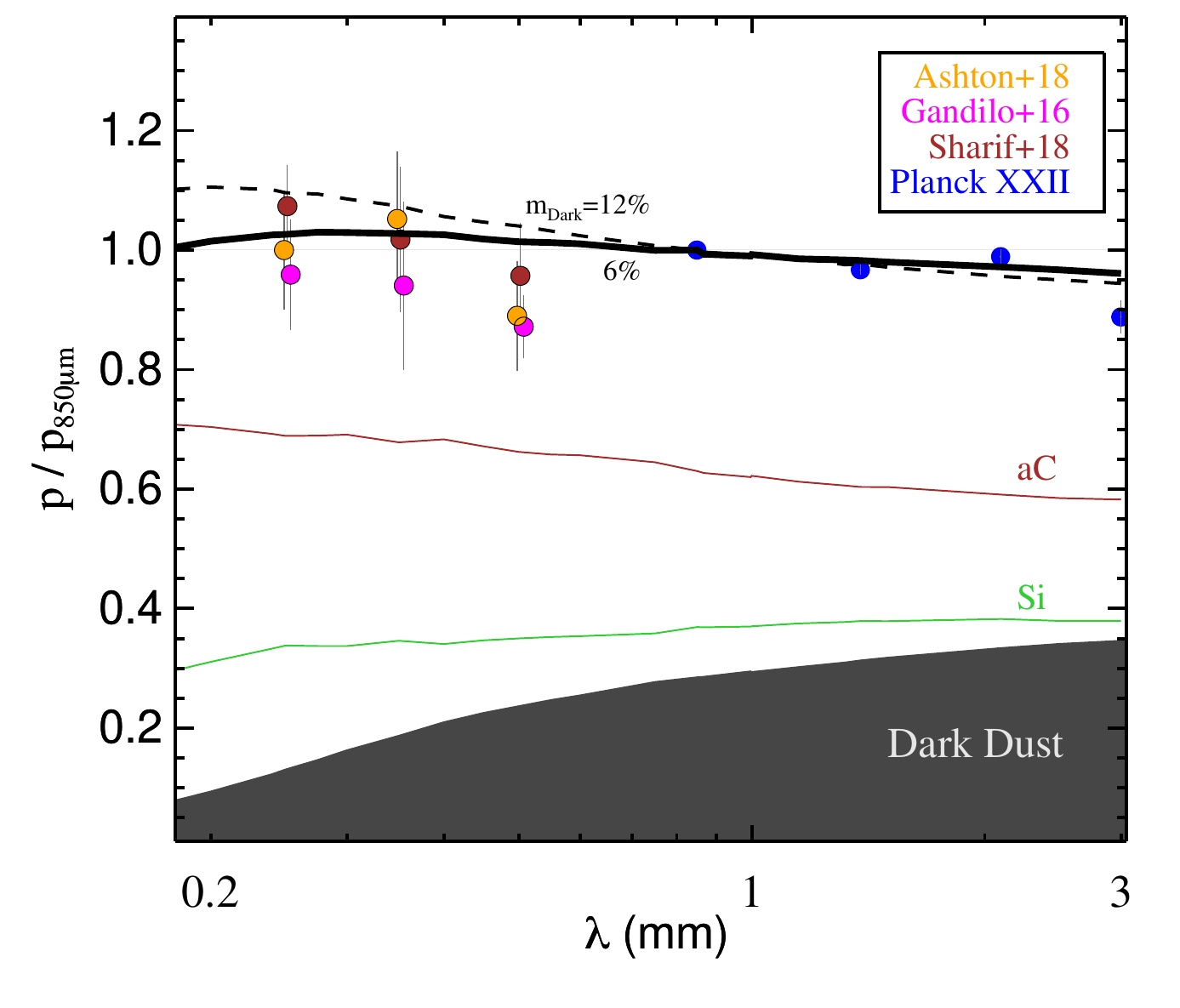}
\caption {Polarized emission of the diffuse ISM between $0.2 - 3$\,mm
  normalized to the polarization at $850\,\mu$m. Observations
  (circles) with $1\sigma$ error bars in blue by \cite{Planck15,
  Planck20}, in magenta by \cite{Gandilo16}, in brown
  by \cite{Shariff19}, in orange by \cite{Ashton18}; as tabulated
  by \citet{HD21}. Model with $r_{\rm {Dark}}^{+} = 1 \, \mu$m
  (Table~\ref{para.tab}) and $m_{\rm {Dark}}=6$\,\% of mass in dark
  dust (full line), the contributions (dashed) from large amorphous
  carbon (brown) and large silicate grains (green) as well as the
  total polarization fraction adopting $m_{\rm {Dark}}=12$\,\%
  (dashed) is shown. The contribution from dark dust when treated as
  prolate particles are indicated by the area in
  grey. \label{subpol.fig} }
\end{figure}

%%%%%%%%%%%%%%%%%%%%%%%%%%%%%%%%%%%%%%%%%%%%%%%%%%%%%

The Planck bandpasses at 0.85\,mm and 3\,mm includes dust and other
emission components such as the CO~$(3-2)$ or CO~$(1-2)$ line
transitions, and at longer wavelengths free--free or synchrotron
radiation \citep{Galametz14}. The two least contaminated bandpasses
near 1.4\,mm and 2.1\,mm are used to estimate a lower limit of the
amount of dark dust. The mass of dark dust is varied in the $r_{\rm
{Dark}}^{+} \ga 0.75\,\mu$m models until a best fit is
found. Typically the mass in dark dust is at least $m_{\rm {Dark}} \ga
6$\,\% of the total dust mass (Table~\ref{para.tab}, Col.~3) to fit
the Planck observed emission.

The polarized dust emission at $\lambda \simgreat 0.85\,\mu$m is
observed by \citet{Planck15, Planck20} and between $0.25 \simgreat
\lambda \simgreat 0.85 \,\mu$m for selected areas on the sky by
BLASTPol \citep{Gandilo16, Ashton18, Shariff19}. The polarization
spectrum normalised to the fractional polarization at 0.85\,mm is
tabulated by \citet{HD21} and is shown in Fig.~\ref{subpol.fig}. It
features an astonishingly flat spectrum that is within 1$\sigma$ of
11\% constant and challenges several dust models \citep{DH21}. The
best fit of the $r_{\rm {Dark}}^{+} = 1\,\mu$m dark dust model to the
polarized emission spectrum is shown in Fig.~\ref{subpol.fig} using
parameters of Table~\ref{para.tab}. The total fractional FIR/mm
polarization spectrum of the model is within $\sim 5$\,\% constant and
aC grains are the dominating contributor over the Si grains.

Dark dust when treated as prolate particles and alignment efficiency
as aC grains give in the FIR a contribution of $\la 10$\,\% and in the
mm range $\sim 35$\,\% to the total polarization. The flatness of the
submm/mm polarization spectrum provides an upper limit of the total
amount of dark dust. This is shown in Fig.~\ref{sed.fig} where the
polarization in the FIR is overestimated when increasing the dark dust
mass to $m_{\rm {Dark}} \ga 12$\,\% of the total dust mass.

The consistency of the models between optical and submm polarization
is verified. Polarized extinction and polarized emission are tightly
connected when produced by the same grains. For the diffuse ISM, a
characteristic value is provided by \citet{Planck20} and is given in
Eq.~\ref{Pratio.eq}. The dark dust models cope with this
optical-to-submm polarization ratio by adopting an optical depth of
$1.3 \la \tau_{\rm V} \la 1.37$ (Col.~18, Table~\ref{para.tab}). This
is somewhat smaller than derived for sightlines with translucent
clouds by \citet{Guillet18}, which have a slightly larger reddening
than the diffuse ISM. The optical depth is converted to a dust
reddening using the model-derived total-to-selective extinction
$R_{\rm V}$ (Eq.~\ref{Rv.eq}). The reddening of the models is
${E(\rm{B} - \rm{V})} = \tau_{\rm V}/1.086/R_{\rm V}$ (Col.~17-18 of
Table~\ref{para.tab}) and range between $0.40 - 0.42$\,mag, which is
in good agreement with the observationally derived mean of 0.45\,mag
\citep{FM07}.

%%%%%%%%%%%%%%%%%%%%%%%%%%%%%%%%%%%%%%%%%%%%%%%%%%%%%%%%%%%%%%%%%%%%%%

%%%%%%%%%%%%%%%%%%%%%%%%%%%%%%%%%%%%%%%%%%%%%%%%%%%%%%%%%%%%%%%%

\begin{table*}[!htb]
\scriptsize
\begin{center}
\caption {Parameters of the dust model for the
diffuse ISM using \citet{Demyk22} optical constants for amorphous
silicate grains. \label{d.tab}}
\begin{tabular} {l r c c | c c r r | c c c r r | c c c c c c}
\hline\hline
1 & 2 & 3 & 4 & 5 & 6 & 7 & 8 & 9 & 10 & 11 & 12 & 13 & 14 & 15 & 16 & 17 & 18 \\
\hline
& & & & & & & & & & & & & & & & & \\
\multicolumn{4}{c|}{Sample} & \multicolumn{4}{c|}{Abundances$^{\dagger} \frac{[\rm{X}]}{[\rm{H}]}$} & \multicolumn{5}{c|}{Sizes } & & \multicolumn{4}{c}{Derived quantities$^{\ddagger}$ } \\
& & & & & & & & & & & & & & & & & \\ 
\hline
ID &
Composition &
$\mu$ &
$\rho$ &
%
%$\frac{[\rm{C}]}{[\rm{H}]}_{\rm {aC}}$ &
%$\frac{[\rm{C}]}{[\rm{H}]}_{\rm {VSG}}$ &
%$\frac{[\rm{C}]}{[\rm{H}]}_{\rm {PAH}}$ & 
vSi & aC & VSG & PAH &
$q$ &
$r_{\rm {Si}}^{+}$ &
$r_{\rm {ac}}^{+}$ &
$r_{\rm {pol, Si}}^{-}$ &
$r_{\rm {pol, aC}}^{-}$ &
$\chi^2$ &
$\frac{[\rm{C}]}{[\rm{Si}]}$ &
$\frac{M_{\rm {gas}}}{M_{\rm {dust}}}$ &
$R_{\rm {V}}$ &
$\tau_{\rm {V}}$ \\
& & ($u$) & (g/cm$^{3})$& & \multicolumn{3}{c|}{(ppm)} & & \multicolumn{4}{c|}{(nm)} & & & & \\
\hline
X35\/ & 0.65 MgO$-$0.35 SiO$_2$ & 141 & 2.7 & 3 & 72 & 11 & 9 & 3.2 & 261 & 262 & 92 & 222 & 1.7 & 5.0 & $<115$ & $-$ & $-$ \\
X40\/ & 0.60 MgO$-$0.40 SiO$_2$ & 121 & 2.7 & 4 & 29 & 7 & 8 & 2.9 & 286 & 261 & 72 & 88 & 0.5 & 2.4 & $<168$ & 2.94 & $-$ \\
X50A & 0.50 MgO$-$0.50 SiO$_2$ & 100 & 2.7 & 3 & 57 & 10 & 6 & 3.0 & 267 & 248 & 54 & 72 & 0.8 & 4.0 & $<129$ & 3.10 & 1.04 \\
X50B & 0.50 MgO$-$0.50 SiO$_2$ & 100 & 2.7 & 4 & 63 & 11 & 7 & 3.0 & 257 & 251 &102 & 233 & 0.8 & 4.3 & $<162$ & 3.14 & $-$ \\
E10\/ & Mg$_{0.9}$Fe$^{3+}_{0.1}$ SiO$_3$ & 99.9 & 2.8 & 5 & 77 & 12 & 8 & 2.9 & 246 & 230 & 59 & 84 & 0.7 & 4.8 & $<148$ & 3.08 & $-$ \\
E20\/ & Mg$_{0.8}$Fe$^{3+}_{0.2}$ SiO$_3$ & 99.3 & 2.9 & 3 & 41 & 10 & 6 & 3.0 & 266 & 253 & 72 & 97 & 0.7 & 3.1 & $<141$ & 3.09 & $-$ \\
E30\/ & Mg$_{0.7}$Fe$^{3+}_{0.3}$ SiO$_3$ & 99.8 & 3.0 & 5 & 76 & 13 & 9 & 3.1 & 259 & 227 & 84 & 183 & 0.5 & 4.9 & $<165$ & 2.99 & $-$ \\
E40\/ & Mg$_{0.6}$Fe$^{3+}_{0.4}$ SiO$_3$ & 99.3 & 3.1 & 5 & 75 & 12 & 9 & 3.1 & 235 & 252 & 57 & 92 & 0.6 & 4.9 & $<161$ & 3.05 & 1.15 \\
E10R & Mg$_{0.9}$Fe$^{2+}_{0.1}$ SiO$_3$ & 99.9 & 2.8 & 4 & 64 & 10 & 8 & 3.0 & 257 & 238 & 66 & 92 & 0.7 & 4.2 & $<159$ & 3.05 & $-$ \\
E20R & Mg$_{0.8}$Fe$^{2+}_{0.2}$ SiO$_3$ & 99.3 & 2.9 & 3 & 46 & 9 & 6 & 3.0 & 267 & 260 & 69 & 92 & 0.7 & 3.3 & $<160$ & 3.11 & $-$ \\
E30R & Mg$_{0.7}$Fe$^{2+}_{0.3}$ SiO$_3$ & 99.8 & 3.0 & 6 & 85 & 13 &10 & 3.0 & 235 & 234 & 88 & 192 & 0.5 & 5.2 & $<159$ & 3.00 & 1.18 \\
E40R & Mg$_{0.6}$Fe$^{2+}_{0.4}$ SiO$_3$ & 99.3 & 3.1 & 5 & 73 & 13 & 8 & 3.2 & 237 & 244 & 59 & 102 & 0.6 & 4.8 & $<162$ & 2.98 & 1.14 \\
\hline
\end{tabular}
\end{center}
{\bf {Notes.}} Notation as in Table~\ref{para.tab}. $^{\dagger}$
Abundance of large amorphous silicate grains are [Si]/[H] =
15\,ppm. $^{\ddagger}$ Models that do not fit the reddening or
polarised emission \citep{Planck20} are indicated by symbol '$-$'
and models that do not fit the dust emission \citep{Dwek97,
Planck15} are indicated by symbol '$<$'.

\end{table*}

%%%%%%%%%%%%%%%%%%%%%%%%%%%%%%%%%%%%%%%%%%%%%%%%%%%%%%%%%%%%%%%%

\begin{figure*}[htb!]
\begin{center}
\includegraphics[width=9.cm,clip=true,trim=4.5cm 6.5cm 3.5cm 5cm]{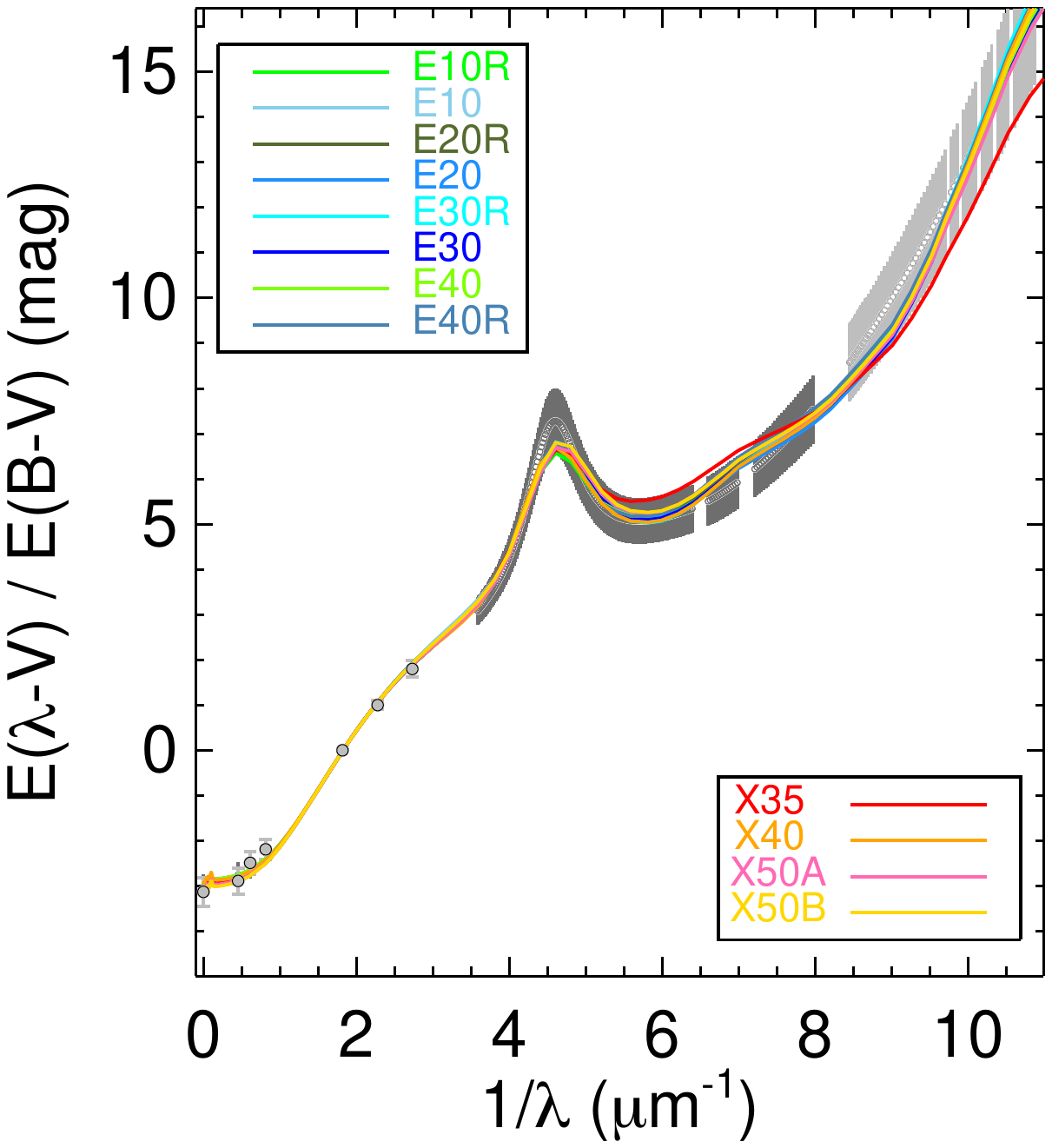}
\includegraphics[width=9.cm,clip=true,trim=4.5cm 6.5cm 3.5cm 5cm]{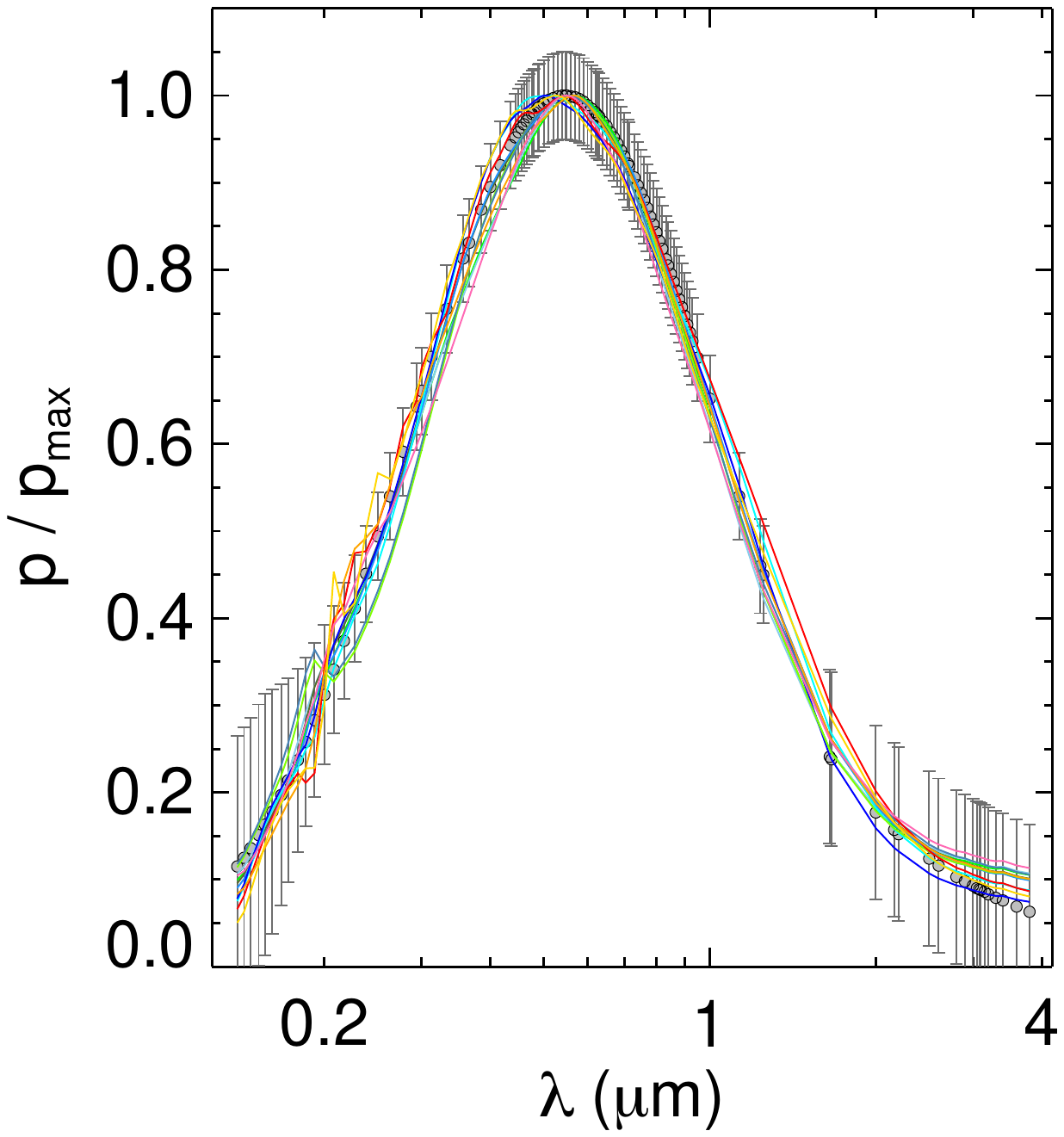}
\end{center}
\caption {Best fits (full line) to the reddening (left) and
starlight polarisation (right) using \citet{Demyk22} optical
constants as labelled. Model parameters as in Table~\ref{d.tab}
and data as in Fig.~\ref{redd.fig} and Fig.~\ref{Serk.fig},
respectively.\label{dr.fig}}
\end{figure*}

%%%%%%%%%%%%%%%%%%%%%%%%%%%%%%%%%%%%%%%%%%%%%%%%%%

\begin{figure*}[htb!]
\begin{center}
\includegraphics[width=14cm,clip=true,trim=4.cm 8.cm 3.5cm 8.5cm]{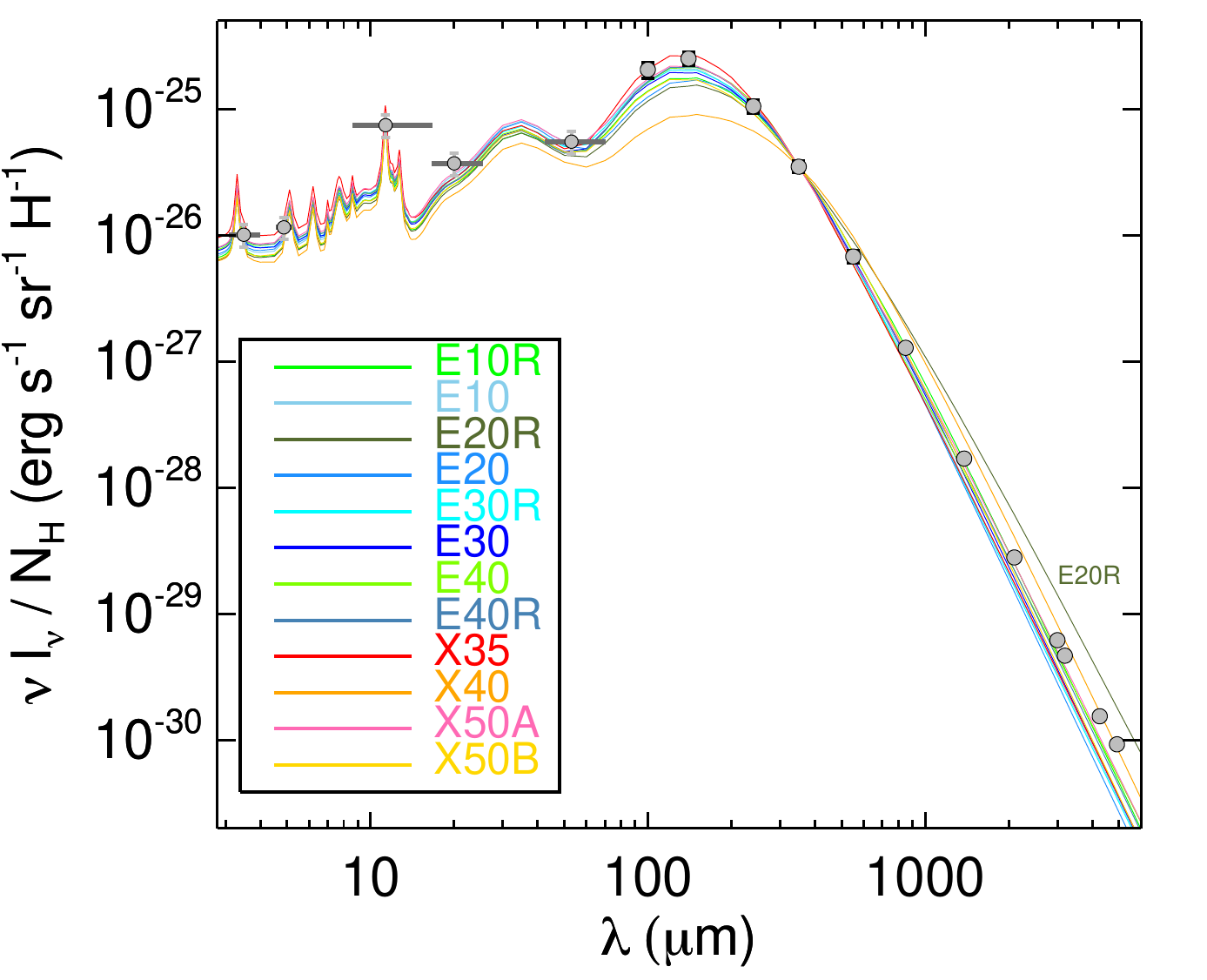}
\end{center}
\caption {Dust emission of the diffuse ISM for models using
\citet{Demyk22} optical constants (as labelled) that fit
simultaneously the reddening and starlight polarization
(Fig.~\ref{dr.fig}). Model parameters as in Table~\ref{d.tab} and
data as in Fig.~\ref{sed.fig}. \label{de.fig}}
\end{figure*}

%%%%%%%%%%%%%%%%%%%%%%%

\section{Laboratory studies of cosmic dust analogous\label{lab.sec}}

\begin{figure*}
\begin{center}
\includegraphics[width=9.1cm,clip=true,trim=4.cm 6.5cm 3.5cm 5cm]{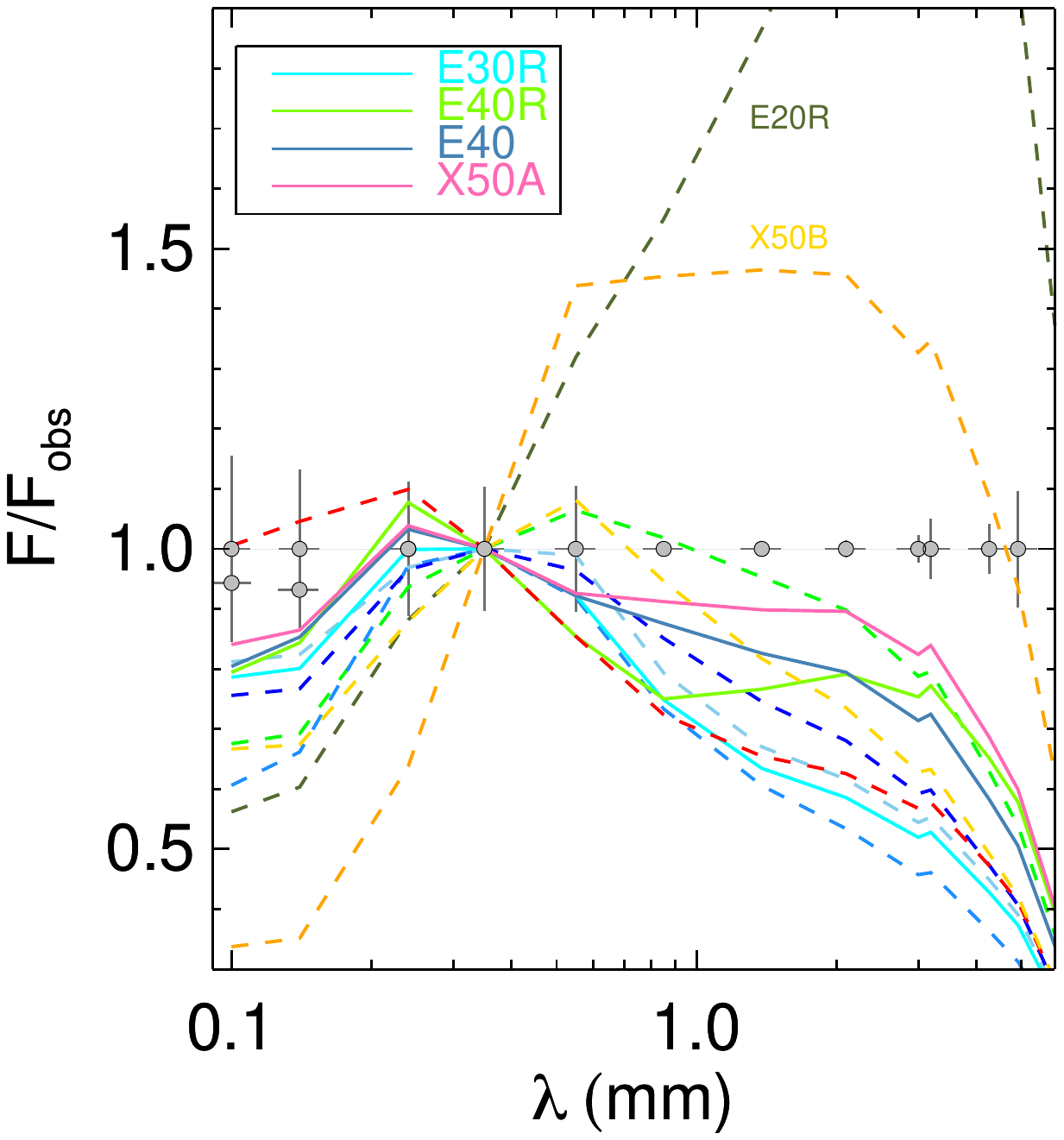}
\includegraphics[width=9.1cm,clip=true,trim=4.cm 6.5cm 3.5cm 5cm]{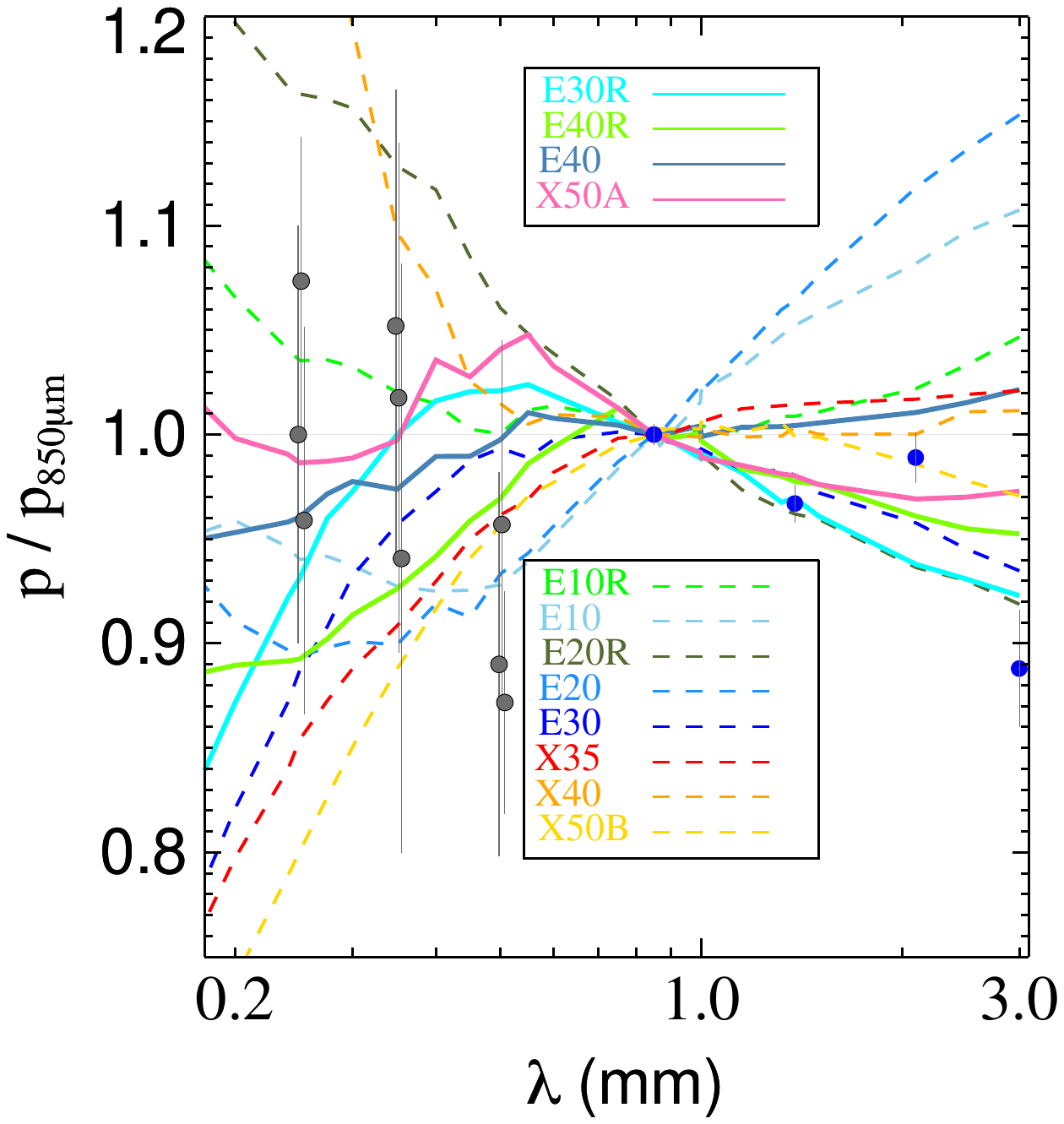}
\end{center}
\caption {Flux ratio to the observed photometry (left) and the
polarised emission spectrum (right) with data as in
Fig.~\ref{sed.fig} and Fig.~\ref{subpol.fig},
respectively. Models using \citet{Demyk22} optical constants (as
labelled) that do not fit the polarised emission spectrum are
shown by dashed lines otherwise by a full line. \label{dfp.fig}}
\end{figure*}

\begin{figure*}
\begin{center}
\includegraphics[width=9.1cm,clip=true,trim=4.cm 6.5cm 3.5cm 5cm]{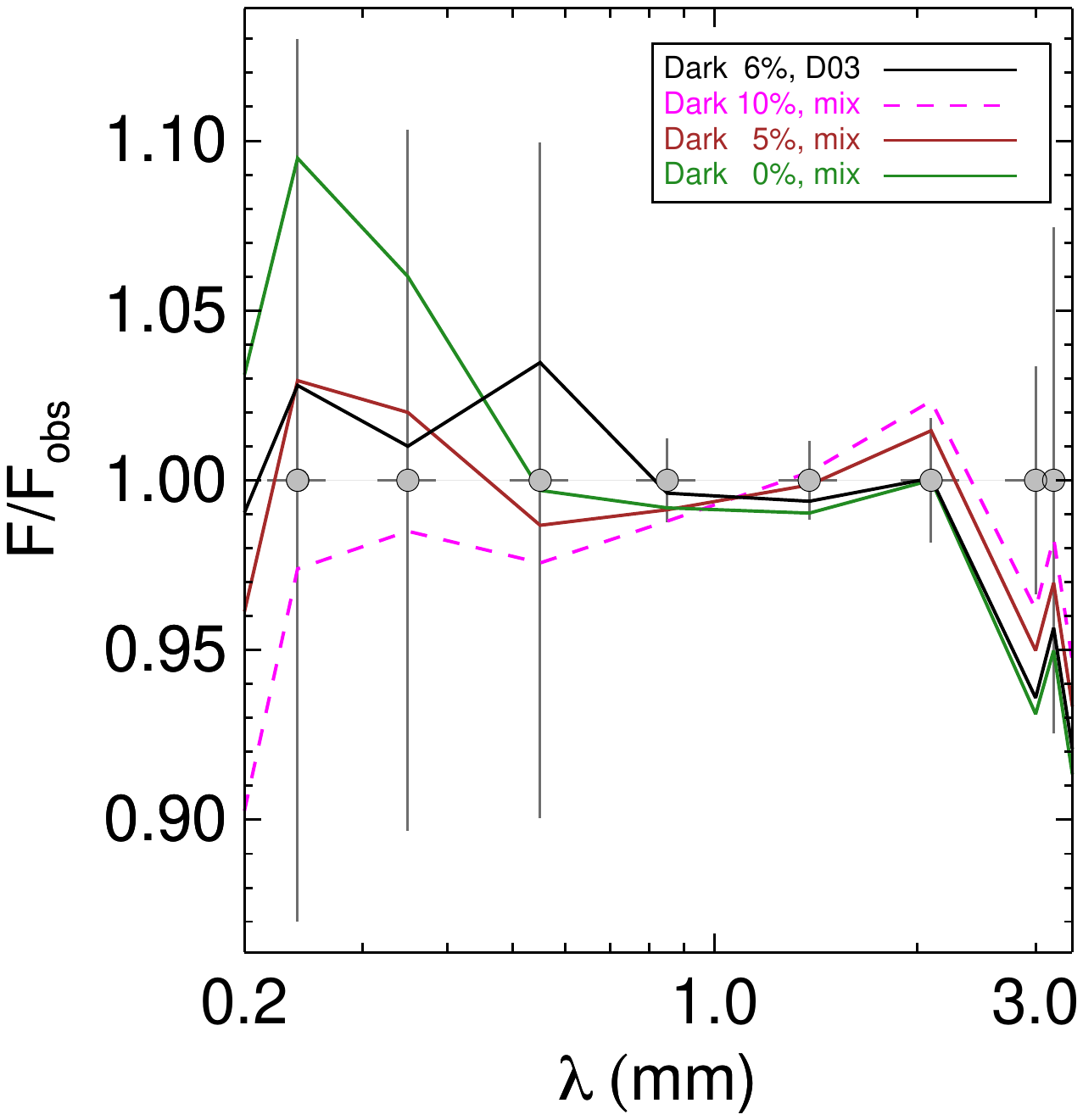}
\includegraphics[width=9.1cm,clip=true,trim=4.cm 6.5cm 3.5cm 5cm]{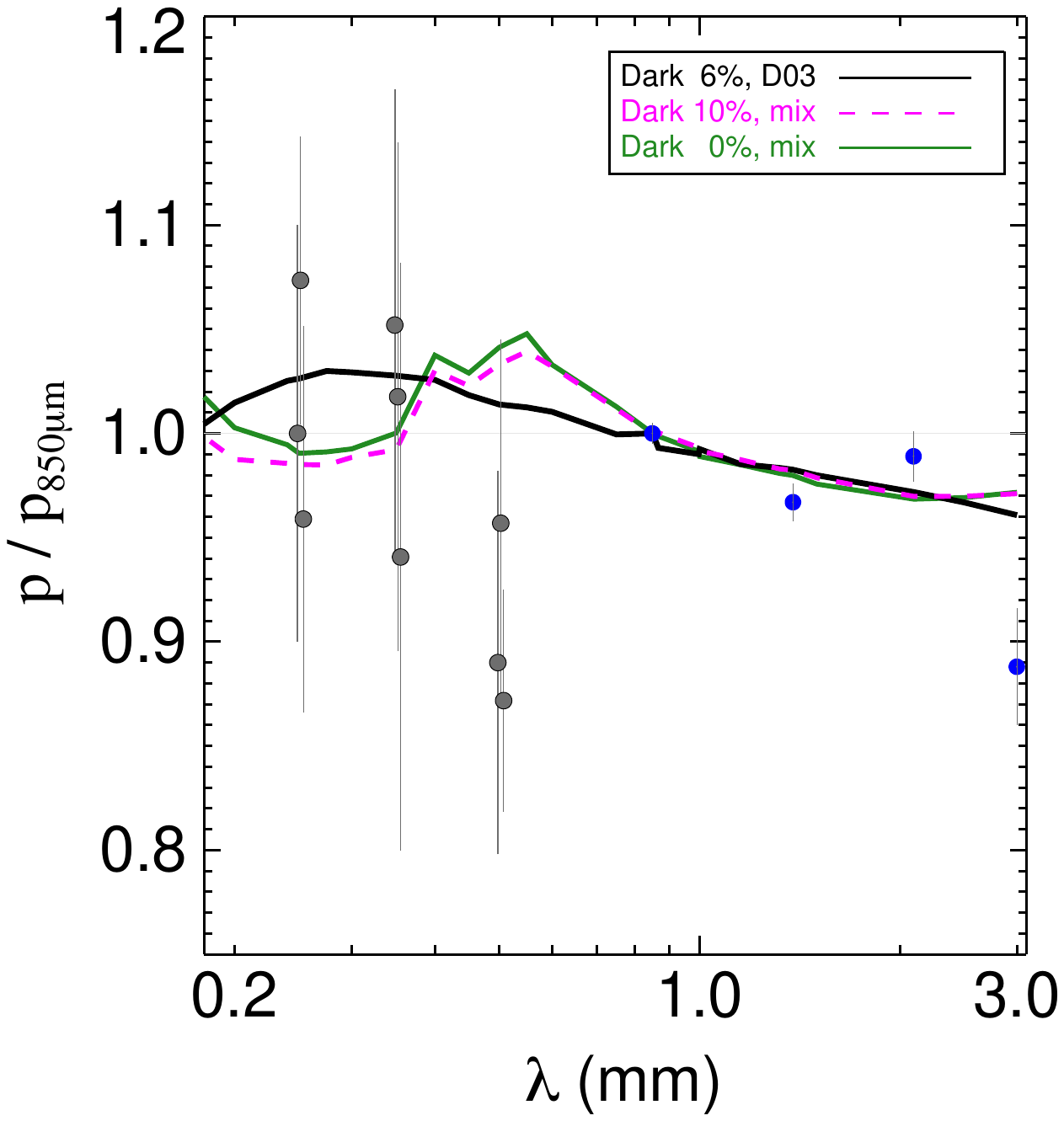}
\end{center}
\caption {As Fig.~\ref{dfp.fig} for a 97:3 mix in mass of the
MgO$-$0.5 SiO$_2$ (X50A) and Mg$_{0.8}$Fe$^{2+}_{0.2}$ SiO$_3$
(E20R) \citet{Demyk22} samples. This model with a contribution to
the total dust mass of 0 (green), 5\% (brown), and 10\% (magenta)
of dark dust is shown. The best fit using \citet{Draine03} optical
constants and parameters as in Table~\ref{para.tab} for the
$r_{\rm {Dark}}^{+} = 1 \, \mu$m model (black) is shown for
comparison. \label{dmix.fig}}
\end{figure*}

Laboratory studies of cosmic dust analogous foster our knowledge in
several research fields \citep{Jaeger20}: The experiments enlighten
our understanding of processes leading to the formation or destruction
of dust particles \citep{Jones94}, the growth of grains to pebbles and
planetesimals, and how dust eventually evolves into planets
\citep{Wurm98, Blum08, Wurm21}. In the laboratory, chemical reaction
paths of interstellar gas with charged or uncharged dust and
nanoparticles and their interaction on the grain surfaces can be
simulated \citep{Salama96, Herbst01, Jones21}. Scattering matrices and
phase functions of several irregular shaped particles have been
measured \citep{Munoz20}.

Experimental results of optical constants for a range of dust
materials have been obtained. Although, they are limited in the
wavelength coverage (\citet{Dorschner95, Mennella98, Jaeger03}, see
the Heidelberg - Jena - St.~Petersburg database~{\footnote
  {https://www2.mpia-hd.mpg.de/HJPDOC}}). Models aiming to explain
simultaneously the dust absorption and emission of polarized and
unpolarised light require a consistent set of optical constants from
the Lyman limit to about 1\,cm, at least. In that wavelength range,
optical constants derived from laboratory experiments are available
for various carbon materials. These have been incorporated in dust
models; see \citet{Zubko04} for a comprehensive study. For amorphous
silicate grains, a complete set of laboratory-derived optical
constants suited for dust modelling was not available until the work
by \citet{Demyk22}. Commonly, and as used in Sect.~\ref{fitting.sec},
the semi-empirical set by \citet{Draine03} is applied. These optical
constants are based on laboratory measurements in the UV/optical by
\citet{Huffman73}, on observations in the NIR and MIR, and some
extrapolation by a power-law to longer wavelengths
\citep{DraineLee84}.  The same extrapolation is used by
\citet{Jones17}, who considered in the UV/optical range optical
constants of amorphous silicate dust by \citet{Scott96} and added
metallic Fe and FeS inclusions to reproduce NIR observations. The
exponent of the power-law was modified by \cite{DH21} to accommodate
the deficit submm/mm emission that was present in all of the previous
dust models when confronted with the observations by the Planck
mission \citep{Ysard20}.

Recently, \citet{Demyk22} calculated optical constants between
$10^{-2} - 10^{5}\,\mu$m of four Mg-rich glassy silicate dust
particles with stoichiometry from about enstatite to olivine and eight
samples of Mg- and Fe-rich silicates with stoichiometry close to
pyroxene. For the samples, the mass absorption coefficients were
measured at temperatures between $10 - 300$\,K and between 5 to $\sim
1000\,\mu$m \citep{Demyk17b,Demyk17}. The optical constants of the
samples show at temperatures above 30\,K and wavelengths $\simgreat \,
80 \, \mu$m a significant temperature dependence. The dust absorption
cross section at 300\,K is increased by about one order of magnitude
when compared to the sample measured at 30\,K. This temperature
dependence of the cross-section has so far been neglected in dust
models. It results in an overestimate of the derived dust masses that
is important for environments where dust is heated to such high
temperatures \citep{Fanciullo20}. The optical constants remain
constant between 10\,K and 30\,K. In the diffuse ISM of the Milky Way
the temperatures of large grains stay below $\sim
20$\,K. \citet{Demyk22} adopted in the wavelength range between 0.5 to
$1 \, \mu$m, the refractive indices as measured for similar kinds of
amorphous silicate grains by
\citet{Dorschner95} and \citet{Jaeger03}. The extrapolation of the
measurements to the entire wavelength range is a delicate issue.
Besides the numerical challenges, detailed knowledge of the sample is
required such as the bulk density of the materials, the grain shapes
and structures, as well as the particle size distribution of the
agglomerates. These parameters were derived by investigating images of
the samples obtained by using Transmission Electron Microscopy
\citep{Demyk22}. Samples labelled 'E' consist of submicron-sized
particles with irregular shapes that are characterised by prolates
with axial ratios of $a/b \sim 2$ and glassy silicate with $a/b \sim
1.5$. These axial ratios are used in the following.

The fitting procedure of Sect.~\ref{fitting.sec} is applied to the
dust models by replacing the optical constants of the large silicate
grains with those by \citet{Demyk22} measured at 10\,K. The alignment
efficiency of silicates is set to $\delta_0 = 1 \ \mu$m. First,
the contribution of dark dust is neglected. The models that are
consistent with the abundance constraints (Eq.~\ref{abu.eq}) and that
best fit the reddening and starlight polarisation and the emission of
polarised and unpolarised light simultaneously are shown in
Fig.~\ref{dr.fig} - Fig.~\ref{dfp.fig}. Sample characteristics, dust
parameters, and derived quantities of these models are summarised in
Table~\ref{d.tab}. In that table, columns~1 - 4 provide the
identifier, composition, molecular weight $\mu$ of the mean
composition, and bulk density $\rho$ of the sample as given by
\citet{Demyk22}. Parameters for the abundances of the dust populations
(columns~4 - 8) and their size parameters (columns~9 - 13) are
specified. The derived quantities such as the dust abundance ratio
(col.~15), the dust-to-gas mass ratio (col.~16), the
total-to-selective extinction (col.~17), and the optical depth
(col.~18) that match the optical-to-submm polarization ratio of
${p_{850\,\mu{\rm m}}} / {(p_{\rm {V}}/\tau_{\rm {V}})} = 4.31$
(Eq.~\ref{Pratio.eq}) by \citet{Planck20}. The goodness parameter of
the fit is given in col.~14.

The mean reddening curve is fit by all samples reasonably well except
for sample X35, which overestimates the reddening between {$5 -
7$\,($\mu$m$^{-1}$)} and underestimates the far UV rise
(Fig.~\ref{dr.fig}, left). The starlight polarisation spectrum is fit
within the errors by all models (Fig.~\ref{dr.fig}, right), although
at somewhat larger dispersion than the dark dust model shown in
Fig.~\ref{Serk.fig}.

However, none of these models fit the dust emission
(Fig.~\ref{de.fig}). All models except sample X35 underestimate the
FIR and all models except samples E20R and X50B underestimate the
emission $>0.5$\,mm (Fig.~\ref{de.fig}). Thus estimates of the
dust-to-gas mass ratios for these models shall be taken with
caution. The model features of the dust emission become more apparent
in the left panel of Fig.~\ref{dfp.fig}, where the ratio of the flux
to the observed photometry is shown. Samples that do not fit the
observed polarised emission spectrum are indicated by a dashed line
and those that do are shown by full lines, those are samples E30R,
E40R, E40 and X40. The potpourri of the fits to the polarised emission
spectrum is shown in the right panel of Fig.~\ref{dfp.fig}. Samples
that do not fit the $0.25 - 0.5$\,mm polarisation spectrum by
\cite{Gandilo16}, \cite{Shariff19}, and \cite{Ashton18} within 15\% or
deviates by more than $5 \sigma$ from the \citet{Planck20}
polarisation spectrum in the millimetre are indicated by dashed lines.

Mixing the curves that are shown in the panels of Fig.~\ref{dfp.fig}
allows finding a flattened curve that better fits the data. The sample
X50A appears particular attractive as it fits the polarization data
and the FIR/submm emission up to 0.8\,mm already. That model only
falls short in the mm-region and this deficit emission may be
filled-up using sample E20R that gravely overestimates the submm/mm
emission. Indeed a 97:3 mix in mass of the MgO$-$0.5 SiO$_2$ (X50A)
and Mg$_{0.8}$Fe$^{2+}_{0.2}$ SiO$_3$ (E20R) provides a simultaneous
fit to all the observing constraints. This dust mix without a
contribution by dark dust is shown in Fig.~\ref{dmix.fig} together
will models that include dark dust contributing to 5\% and 10\% to the
total dust mass, respectively. The best fit using \citet{Draine03}
optical constants and parameters as in Table~\ref{para.tab} for the
$r_{\rm {Dark}}^{+} = 1 \, \mu$m model is shown for comparison. All
four models provide a similarly good fit to the data.

%%%%%%%%%%%%%%%%%%%%%%%%%%%%%%%%%%%%%%%%%%%%%%%%%%

\section{Conclusion \label{conclusion.sec}}

Dark dust was suggested for the unification of spectroscopic and
parallax-derived distances. Such grains have a wavelength-independent
reddening and non-selective extinction in the optical were already
considered by \cite{Trumpler}. The extinction properties of these
particles require that they are large. Their absorbed energy will be
re-emitted in the FIR-mm.

In this paper dark dust has been included as an additional grain
population. The model copes with the current state of observations and
provides constraints on the physical properties of dust in the diffuse
ISM. Major observations that have been verified are representative
solid-phase element abundances, far UV-NIR reddening, optical-NIR
polarized extinction, FIR-mm dust emission of polarized and
unpolarized light, and the ratio of the optical-to-submm polarization.
A dust model with 11 parameters for the various grain abundances,
sizes, and alignment properties accounts for these data
simultaneously. The parameter space is explored by a vectorised
iterative fitting procedure to derive the physical properties of dark
dust. Different sets of optical constants for amorphous silicate
grains are analysed. The principal findings of that study are:

\begin{enumerate}
\item A large set of models that account for the mean reddening,
extinction and starlight polarization curves of the Milky Way fail
to respect the adopted abundance constraints.

\item A detailed fit to the Serkowski curve is derived assuming
  different alignment characteristics of large aC and Si grains.

\item Dark dust consists of micrometre-sized particles of at least
$r_{\rm {Dark}}^{+} \simgreat 1 \, \mu$m.

\item Dark dust provides a significant wavelength-independent
reddening in the UV/optical and up to wavelengths $\lambda \la
r_{\rm {Dark}}^{+}$.

\item In the FIR/mm the extinction cross-section of dark dust is of
  similar strengths than Si grains and shows a slightly less steep
  slope than large grains.

\item The detected Planck excess emission at $0.8 - 3$\,mm that
  previous models cannot explain without ad-hoc changes in the slope
  of the dust emissivity or adjustments of the optical constants is
  provided by the emission of very cold ($8 - 12$\,K) dark dust. Very
  cold dust is frequently observed in other non-active galaxies.

\item The observed flatness of the submm/mm polarization spectrum is
held by the model unless the mass in dark dust is $\la 12$\,\% of
the total dust mass.

\item Dark dust, when treated as aligned prolate particles with $r_{\rm
{Dark}}^{+} = 1\,\mu$m, does not provide a significant contribution
to the observed polarized extinction, polarization in the $10\,\mu$m
silicate band, and polarized emission at $\lambda \la 1$\,mm. At
longer wavelengths ($1-3$\,mm) dark dust may contribute to $\sim
1/3$ of the maximum polarization.

\item Dark dust models fit the characteristic value of the
optical-to-submm polarization ratio assuming an optical depth of
$\tau_{\rm V} = 1.25 \pm 0.13$. The total-to-selective extinction
varies between $2.9 \la R_{\rm V} \la 3.2$, which translates to a
mean reddennig of ${E(\rm{B} - \rm{V})} \sim 0.4$\,mag.

\item The gas-to-dust mass ratio of ${M_{\rm {gas}}}/{M_{\rm {dust}}}
= 126 \pm 4$ in the models using optical constants by
\citet{Draine03}. The estimate is consistent when using a 97:3
mix in mass of silicate dust analogous to the samples by
\citet{Demyk22} of MgO$-$0.5 SiO$_2$ and
Mg$_{0.8}$Fe$^{2+}_{0.2}$. The later mixture provides a good fit to
the observing constraints and still accommodates up to 5 - 10\% of
mass in dark dust.

\end{enumerate}

Present spectro-polarization in the MIR demonstrates the ability to
constrain the stoichiometry of silicate grains. More sensitive MIR
polarization spectra are required along sightlines of the diffuse ISM
to constrain the models presented here. Dark dust with icy grain
mantles would foster grain growth and provide a reservoir for the yet
unaccounted O depletion. For the manifestation of such a model ice
absorption features in the diffuse ISM need to be
detected. Observations shall be possible utilizing the high
sensitivity of the JWST spectrometers.

The dark dust model will be applied to individual sightlines to
explain the absolute reddening and polarization towards selected
stars. This requires a unification of distance estimates to the star
either when derived by spectroscopy or from parallax. Dust models that
ignore dark dust do not account for the unification of both distance
estimates. Dark dust causes significant dimming of light that is not
treated in some contemporary research for example by estimating
spectroscopic distances and the calibration of SN~Ia light curves
\citep{Phillips93, Riess96}. This deserves further investigation.

%%%%%%%%%%%%%%%%%%%%%%%%%%%%%%%%%%%%%

\begin{acknowledgements} {I am grateful to
Karen Demyk for granting me early access to optical constants of the
laboratory samples of amorphous silicate grains. I am grateful to
Endrik Kr\"ugel and Rolf Chini for helpful discussions. }
\end{acknowledgements}

\bibliographystyle{aa}
\bibliography{Ref}

\end{document}